\documentclass[iop,revtex4]{emulateapj}
\usepackage{amsmath,amssymb}
\usepackage{graphicx}

\slugcomment{}

\shorttitle{Two-point Correlation and Network Statistics of LAEs at $z \approx 2.67$}
\shortauthors{Hong et al.}

\begin{document}


\title{Statistics of Two-point Correlation and Network Topology for Lyman Alpha Emitters at $z \approx 2.67$}

\author{ 
Sungryong Hong\altaffilmark{1,2}, 
Arjun Dey\altaffilmark{3},
Kyung-Soo Lee\altaffilmark{4},
\'Alvaro A. Orsi\altaffilmark{5},
Karl Gebhardt\altaffilmark{1},
Mark Vogelsberger\altaffilmark{6},
Lars Hernquist\altaffilmark{7},
Rui Xue\altaffilmark{4}, 
Intae Jung\altaffilmark{1}, 
Steven L. Finklestein\altaffilmark{1},  
Sarah Tuttle\altaffilmark{8}, 
and Michael Boylan-Kolchin\altaffilmark{1} 
}

\altaffiltext{1}{Department of Astronomy, The University of Texas at Austin, Austin, TX 78712, USA}
\altaffiltext{2}{School of Physics, Korea Institute for Advanced Study,  
85 Hoegiro, Dongdaemun-gu, Seoul 02445, Korea}
\altaffiltext{3}{National Optical Astronomy Observatory, 950 N. Cherry Ave., Tucson, AZ 85719, USA}
\altaffiltext{4}{Department of Physics and Astronomy, 
Purdue University, 525 Northwestern Avenue, West Lafayette, IN 47907, USA}
\altaffiltext{5}{Centro de Estudios de Fisica del Cosmos de Aragon, Plaza
de San Juan 1, Teruel, 44001, Spain}
\altaffiltext{6}{Department of Physics, Kavli Institute for Astrophysics and Space Research, 
Massachusetts Institute of Technology, Cambridge, Massachusetts 02139, USA}
\altaffiltext{7}{Harvard-Smithsonian Center for Astrophysics, 60 Garden Street, Cambridge, Massachusetts 02138, USA}
\altaffiltext{8}{Department of Astronomy, University of Washington, Box 351580, Seattle, WA 98195, USA}
\begin{abstract}
We investigate the spatial distribution of Lyman alpha emitting galaxies (LAEs) at $z \approx 2.67$, 
selected from the NOAO Deep Wide-Field Survey (NDWFS), 
using two-point statistics and topological diagnostics adopted from network science.  
We measure the clustering length, $r_0 \approx 4 h^{-1}$ Mpc, 
and the bias, ~$b_{LAE} = 2.2^{+0.2}_{-0.1}$. 
Fitting the clustering with halo occupation distribution (HOD) models results 
in two disparate possibilities: (1) where the fraction of central galaxies is 
$<$1\% in halos of mass $>10^{12}$$M_\odot$; and (2) where the fraction is $\approx$20\%. 
We refer to these two scenarios as the ``Dusty Core Scenario'' for Model\#1 
since most of central galaxies in massive halos are dead in Ly$\alpha$ emission, 
and the ``Pristine Core Scenario'' for Model\#2 since the central galaxies are bright in Ly$\alpha$ emission. 
Traditional two-point statistics cannot distinguish between these disparate models given the current data sets. 
To overcome this degeneracy, we generate mock catalogs for each HOD model 
using a high resolution $N$-body simulation 
and adopt a network statistics approach, which provides excellent topological diagnostics for galaxy point distributions. 
We find three topological anomalies from the spatial distribution of observed LAEs, 
which are not reproduced by the HOD mocks. 
We find that Model\#2 matches better all network statistics than Model\#1, 
suggesting that the central galaxies in $> 10^{12} h^{-1} M_\odot$ halos at $z \approx 2.67$  
need to be less dusty to be bright as LAEs, 
potentially implying some replenishing channels of pristine gas 
such as the cold mode accretion. 
\end{abstract}



\section{Introduction}

The modern cosmology is founded on the cosmological principle   
that the Universe is homogeneous and isotropic. 
The remarkably isotropic Cosmic Microwave Background (CMB; Planck Collaboration XVI 2016)   
strongly supports this cosmological axiom and 
implies, further, the existence of inflationary phase in the early Universe, 
which explains why the Universe is so homogeneous and isotropic within the observable horizon 
(Strarobinsky 1980; Bardeen, Steinhardt \& Turner 1983). 

In contrast, the observed distribution of galaxies looks neither homogeneous or isotropic. 
The gap between the remarkably uniform early Universe and richly structured galaxy distribution 
reflects the complex connections between the cosmic matter distribution and observed galaxy point distribution, 
and emphasizes the importance of identifying useful methodologies 
to quantify the inhomogeneous features in galaxy point distributions. 

Statistics of n-point correlations have been major tools 
for quantifying the spatial distribution of galaxies and have found the critical feature 
of Baryon Acoustic Oscillations (BAOs), 
used for constraining the expansion rates of the Universe 
(Eisenstein, Hu \& Tegmark 1998; Seo \& Eisenstein 2003; Eisenstein et al. 2005; Cole et al. 2005). 
Moreover, each galaxy population has its own spatial clustering property 
(i.e., its own bias from the cosmic dark matter distribution), 
which can be used for testing theories of galaxy formation and evolution 
(e.g., Seljak 2000, Berlind \& Weinberg 2002, Ouchi et al. 2010, Orsi \& Angulo 2017).

As alternatives to the successful n-point statistics, various topological diagnostics 
have been introduced, such as Betti numbers, Minkowski functionals, and genus  
(Gott, Weinberg \& Melott 1987; Eriksen et al. 2004; van de Weygaert et al. 2013; Pranav et al. 2017). 
To identify voids and filaments, various methods have been adopted from other fields of science, 
including minimum-spanning trees (MSTs), watersheds, Morse theory, wavelets, 
and smoothed Hessian matrices (e.g., Barrow, Bhavsar \& Sonoda 1985; Sheth et al. 2003; 
Mart\'inez et al. 2005; Arag\'on-Calvo et al. 2007; Colberg 2007; Sousbie et al. 2008; 
Bond, Strauss \& Cen 2010; Lidz et al. 2010; Cautun, van de Weygaert \& Jones 2013). 
While these topological diagnostics have provided important insights 
into the nature of structure in the Universe, this wide but heterogeneous range 
of applied methodologies reflects how difficult it is to find a consistent 
and comprehensive framework for quantifying and measuring 
the topology of the Universe, in contrast to the successful n-point statistics.

To explore a new way to quantify cosmic topologies, 
Hong \& Dey (2015, hereafter HD15) applied the analysis tools 
developed for the study of complex networks 
(e.g., Albert \& Barab\'asi 2002; Newman 2010) 
to the study of the large-scale galaxy distribution. 
The basic idea is to generate a graph (i.e., ÔnetworkÕ) composed 
of vertices (nodes) and edges (links) from a galaxy distribution, 
and then measure network quantities used in graph theory. 

In this paper, we investigate the spatial distribution of Lyman alpha emitters (LAEs) at $z \approx 2.67$, 
selected from the Bo\"otes field of NOAO Deep Wide-Field Survey (NDWFS),  
utilizing both statistics of two-point correlation and network topology. 
In Section 2, we describe our observed LAE sample. 
In Section 3, we present the two-point statistics of our LAE sample 
and related halo properties from the analyses of Halo Occupation Distributions (HODs). 
In Section 4, we present the network statistics and related topological features. 
We summarize and discuss our findings in Section 5.  
We adopt the AB system for all magnitudes (Gunn \& Oke 1975) 
and the cosmological parameters from Planck Collaboration XVI (2014), 
using the built-in presets of {\it Planck13}~ from ASTROPY (Astropy Collaboration et al. 2013); 
$ \Omega_m = 0.307$, $H_0 = 67.8$, and the flat Universe.
The halo catalogs from Small MultiDark Planck simulation 
are also consistent with {\it Planck13} parameters (SMDPL; Klypin et al. 2014). 
We define $h \equiv h_{\rm 100} \equiv H_{\rm 0}/(100~{\rm km~s^{-1}~Mpc^{-1}})$. 
In this cosmology, the physical scale is $8.15$ kpc/arcsec at $z = 2.67$. 



\section{Observations and Reductions}

\begin{figure}[t]
\centering
\includegraphics[height=4.6 in]{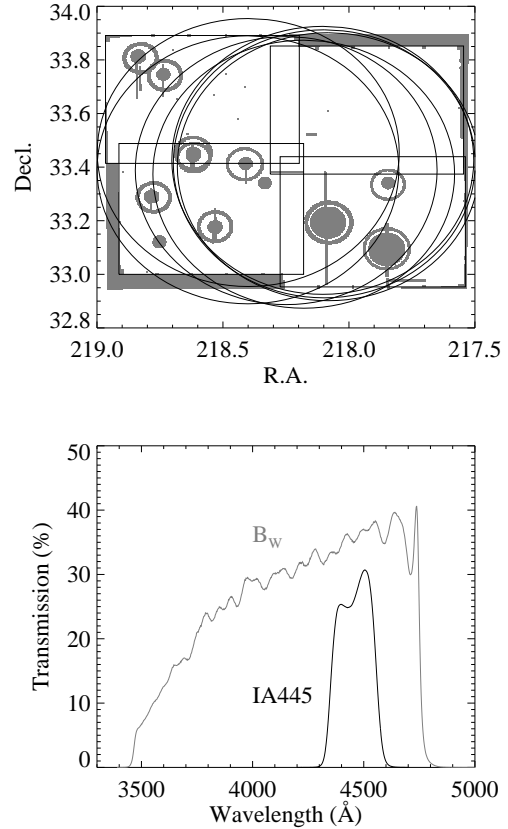}
\caption{The top panel shows our survey area. The photometric masks are shown in grey. 
The open squares represent the 4 pointings of the Subaru/SuprimeCam imaging using the $IA445$ filter. 
The open circles represent the field of views of 7 configurations of MMT/Hectospec observations. 
We trim the margins around the SuprimeCam fields   
by taking the box of R.A. ($\alpha = 218.9^{\circ} - 217.55^{\circ}$) 
and Decl. ($\delta = 32.95^{\circ} - 33.85^{\circ}$). 
The bottom panel shows the filter transmissions of the $IA445$ and $B_W$ filters, 
including all effects from atmosphere, telescope, and optics.  
We choose photometric LAE candidates using the color of these two filters. 
The observed comoving volume covered by the intermediate band survey 
is $82 \times 66 \times 187 ~h^{-3}$Mpc$^3$.
}\label{fig:1}
\end{figure}

We have carried out an intermediate band survey of a ~$\approx1$ square degree area 
in the Bo\"otes field of the NOAO Deep Wide Field Survey (NDWFS; Jannuzi \& Dey 1999) 
aimed at selecting LAEs at $2.55 \lesssim z \lesssim 2.8$. 
We used the $IA445$ filter with the SuprimeCam imaging camera 
on the Subaru telescope to map 4 contiguous fields (Prescott et al. 2008).
The four open squares in the top panel of Figure~\ref{fig:1} show the coverage of our survey, 
and the complex shapes painted in grey represent the observing mask. 
The bottom panel of Figure~\ref{fig:1} shows filter transmission curves for $B_W$ and $IA445$ filters. 
Using Source Extractor (Bertin \& Arnouts 1996), we identify 242,678 objects in this observing field.  
The details about the photometric data can be found in Prescott et al. (2008) and Dey et al. (2016).

\subsection{Candidate Selection}

We define a  sample of LAEs at $z\approx2.67$ using the following photometric criteria:
\begin{eqnarray}
\{ IA445 < 26 \} & \cap & \{(IA445 - B_W) \le -0.5 \} \nonumber \\
 & \cap & \{(IA445 - B_W) \le \frac{4}{9}(26 - IA445) - 0.9\}\nonumber \\
  & \cap & \{(B_W - R) \le 0.8 \}.\label{eq:one}
\end{eqnarray}

\begin{figure*}[t]
\centering
\includegraphics[height=7.0 in]{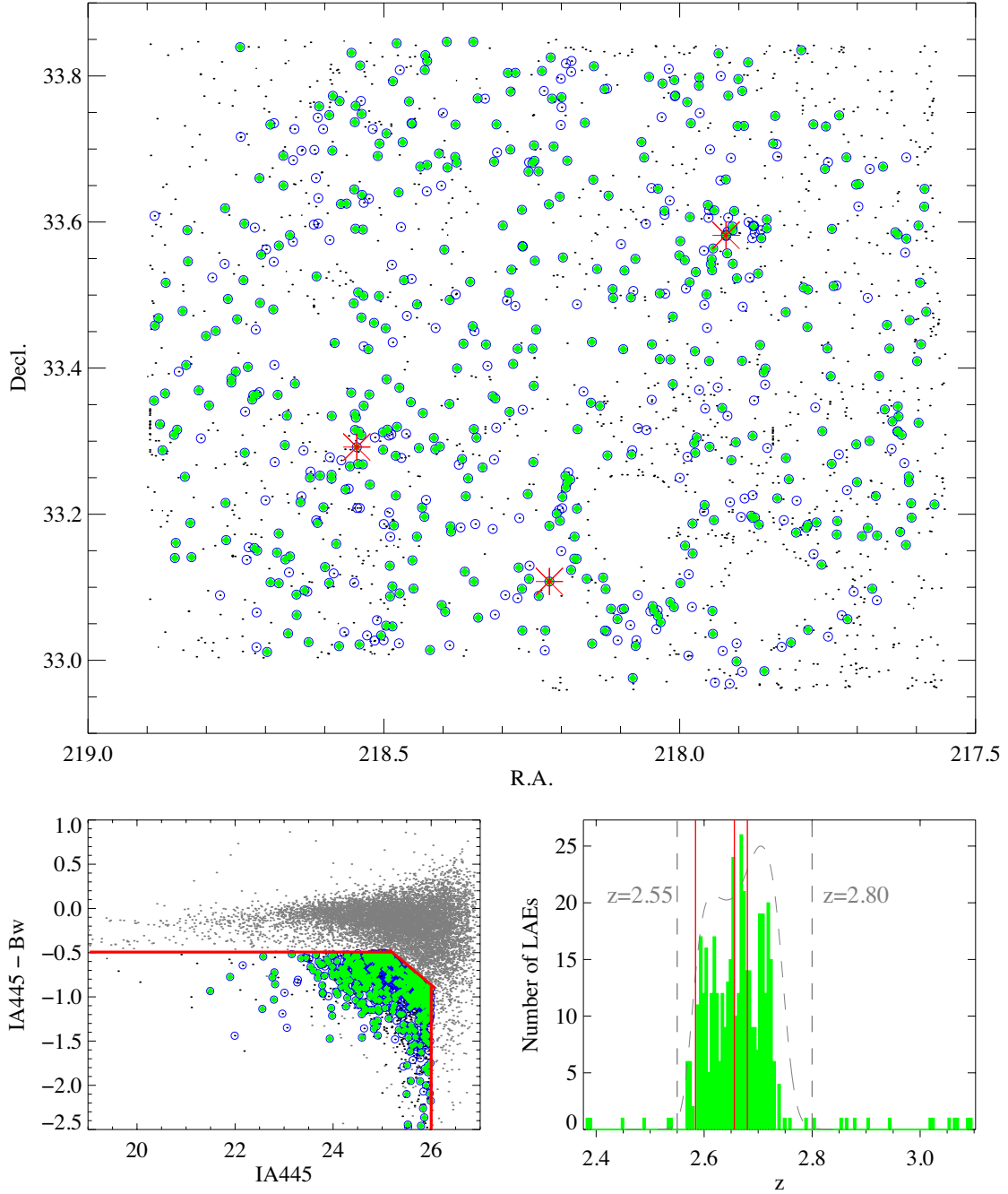}
\caption{ The top panel shows the spatial distribution of 1957 photometric sample pLAE (black dots), 
635 spectroscopically observed targets (blue open circles), and 434 redshift detections (green filled diamonds). 
The big (red) asterisks represent the three Lyman alpha blobs found in our survey field. 
The bottom-left panel shows the color-magnitude distribution, $IA445$ $- ~B_W$ vs. $IA445$, for our source catalog 
(grey dots; $10^4$ objects are plotted from the total 242678 objects) 
and color selection criteria for pLAE (red solid lines).  
The same symbols of black dots, blue open circles, and filled diamonds 
are used for representing the same kinds of objects shown in the top panel. 
The bottom right panel shows the histogram of 434 redshifts (green bars) and 
filter transmission curve for $IA445$ (grey dashed curve). 
The two vertical dashed lines represent the redshifts, $z = 2.55$ and $2.80$, 
respectively and three red solid vertical lines indicate the redshifts 
of three Lyman Alpha Blobs (LABs) discovered in our study.
}\label{fig:2}
\end{figure*}

The bottom-left panel of Figure~\ref{fig:2} shows the color-magnitude diagram 
of $IA445$ $-$ $B_W$ vs. $IA445$, for 10,000 randomly selected sources (grey points) 
from the total 242,678. 
The red solid lines represent our color selection described in Equation~\ref{eq:one}.  
The first two terms in Equation~\ref{eq:one} represent magnitude and color limits. 
The last term of $B_W-R$ color rejects low redshift interlopers. 
From this photometric color selection, we extract 1957 LAE candidates; hereafter, referred to as pLAE.  
We show the spatial distribution of this pLAE sample, using black dots, on the top panel of Figure~\ref{fig:2}. 

\subsection{Spectroscopic redshifts} 

We observed 635 candidates from the total 1957 pLAE sample using Hectospec, 
a multi-object spectrograph on the MMT telescope; the details about the spectroscopic 
data can be found in Hong et al. (2014). 
The seven large open circles in the top panel in Figure~\ref{fig:1} show 
our 7 MMT/Hectospec pointings.  
Within our redshift selection box, $z = 2.55 - 2.80$, 
we confirmed 415 spectroscopic LAEs from the observed 635 pLAE candidates (i.e., a success rate of 65\%). 
We refer to this spectroscopically confirmed subset as zLAE. 
Extrapolating the success rate to the remaining photometric LAE sample using binomial trials, we expect a total sample of $1274\pm17$ zLAEs out of the 1957 pLAE sample.

The bottom-right panel in Figure~\ref{fig:2} shows 
the histogram of redshift detections for zLAE (green bars) and filter transmission curve for $IA445$ (grey dashed curve). 
The two vertical dashed lines represent the lower and upper redshift cutoffs, $z = 2.55$ and $2.80$, 
respectively and three red solid vertical lines indicate the redshifts 
of three Lyman Alpha Blobs (LABs) discovered in our study.  
The top panel shows the spatial distributions of 635 spectroscopically observed targets 
(blue open circles), and 415 zLAE objects (green filled diamonds).
The three red asterisks represent the locations of LABs, 
where their redshifts are $z =  2.680, 2.656,$ and $2.584$ respectively, 
in the order of increasing declination. 



\section{Statistics of Two-point Correlations}

In this section, we investigate the spatial distribution of LAEs using 
 two-point correlation functions by following the conventional clustering studies 
(e.g., Seljak 2000, Berlind \& Weinberg 2002, Roche et al. 2002, Hamana et al. 2004, Zehavi et al. 2004, 
Zheng et al. 2005, Lee et al. 2006, Gawiser et al. 2007, 
Kova\v{c} et al. 2007, Lee et al. 2009, 
Ouchi et al. 2010, Geach et al. 2012,  S\'{a}nchez et al. 2012).

\subsection{Angular Correlation Function $\omega(\theta)$}\label{sec:w}

We measure angular two-point correlation functions for the zLAE and pLAE samples using the estimator 
suggested by Landy \& Szalay (1993; hereafter, the LS estimator),  
\begin{equation}
\omega_{LS}(\theta) = \frac{DD - 2DR + RR}{RR}, \label{eq:ls}
\end{equation}
where DD is the pair count of observed sample, RR of random sample, 
and DR between observed and random samples, within the angular bin $(\theta - \delta\theta/2, \theta + \delta\theta/2)$.  
Since this estimator has been widely used, we only provide a brief description about this method. 
The details can be found in the papers cited above. 

\begin{figure}[t]
\centering
\includegraphics[height=3.5 in]{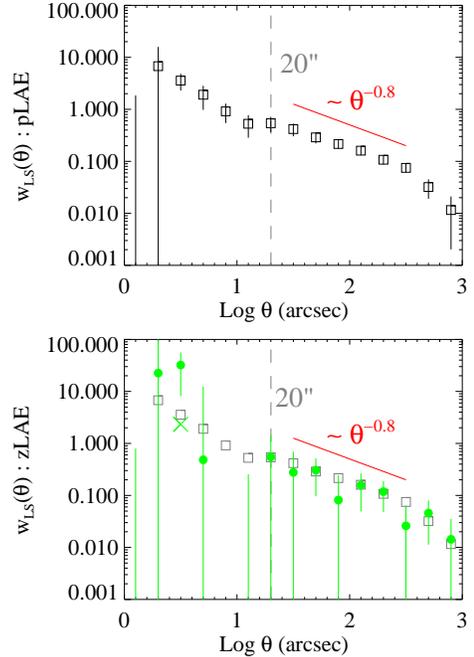}
\caption{The measured angular correlation functions for pLAE (black open squares; top) 
and zLAE (green solid circles; bottom), obtained using the LS estimator. 
The error bars are calculated from bootstrap resampling and represented by vertical lines. 
On the bottom panel, we also add the angular correlation function of pLAE using grey open squares 
for comparison with zLAE. 
Though the uncertainty of zLAE is large, 
both angular correlation functions are consistent, which implies that 
the final LAE sample with complete spectroscopic redshifts 
will not be much different from zLAE and pLAE.    
}\label{fig:3}
\end{figure}

Figure~\ref{fig:3} shows the measured angular correlation functions using the LS estimator for pLAE (black open squares; top) 
and zLAE (green solid circles; bottom). The error bars are calculated from bootstrap resampling (Ouchi et al. 2010). 
On the bottom panel, we also add the angular correlation function 
of pLAE using grey open squares for comparison with zLAE. 
The uncertainty of zLAE is larger due to the small sample size. 
In particular, the small scale clustering at $< 20$\arcsec~ is determined by a small number of close pairs. 
The green `x' mark represents the two-point statistic when we remove the 3 LAEs from zLAEs near 
the largest LAB, LABd05, where we oversample in spectroscopy for investigating 
the environmental effect between LAB and LAEs. Overall, though the uncertainty of zLAE is large, 
the two angular correlation functions are consistent with each other.  

The angular correlation function of the LAEs shows an inflection point 
at scales of 20\arcsec, corresponding to a comoving scale of $0.41 h^{-1}$ Mpc. 
This distinct feature is predicted by halo occupation models, 
where it results from the transition from multiple galaxies occupying 
common halos to each galaxy occupying a single halo.

The LS estimator, $w_{LS} (\theta)$, in Equation~\ref{eq:ls} is 
a normalized quantity of its true angular correlation, $\omega(\theta)$, as   
\begin{eqnarray}
1 + \omega_{LS}(\theta) &=& \frac{1 + \omega(\theta)}{1+\omega_\Omega}, \label{eq:ha} \\
\omega_\Omega &\equiv& \frac{1}{\Omega^2} \int d\Omega_1 d\Omega_2 \omega(\theta), \label{eq:hb} 
\end{eqnarray}
where $\omega_\Omega$ is called ``integral constant'' (hereafter, IC). 
To retrieve the true angular correlation, $\omega(\theta)$, from our measured LS estimator, $\omega_{LS}(\theta)$, 
we need a method to correct this integral constant, $\omega_\Omega$. 
To estimate this IC, we rewrite Equation \ref{eq:ha} and \ref{eq:hb} in more practical forms as 
\begin{eqnarray}
\omega_{LS}(\theta) &=& \frac{\omega(\theta) - \omega_\Omega}{1+\omega_\Omega},\label{eq:hc}\\
\omega_\Omega &\approx& \frac{\sum RR ~ \omega(\theta)}{\sum RR}, \label{eq:hd}
\end{eqnarray}
where Equation  \ref{eq:hc} is rewritten from Equation  \ref{eq:ha} and 
Equation  \ref{eq:hd} is a Monte Carlo integration of Equation  \ref{eq:hb} 
using the same random pairs, RR, in Equation  \ref{eq:ls} (Roche et al. 2002).

\subsection{Interpretations from Single Power-law Correlation Functions}

\subsubsection{Integral Constraint and Self-consistent Fit}\label{sec:poweric}

Unfortunately, we cannot solve Equation~\ref{eq:hc} and \ref{eq:hd}, 
since  $\omega(\theta)$ and $\omega_\Omega$ are coupled, 
and it is the LS estimator, $\omega_{LS}(\theta)$, what we can actually measure from galaxy distribution, 
not $\omega(\theta)$.   
We can resolve this coupling issue if we have some specific constraints on $\omega(\theta)$. 

Conventionally, $\omega (\theta)$ has been assumed to follow a single power-law. 
In this case, we can solve the coupled equations as follows. 
First, we write down the equations  as, 
\begin{eqnarray}
\omega(\theta) &=& A_\omega~ \theta^{-\beta}, \\
\omega_\Omega &\approx& \frac{\sum RR ~A_\omega~\theta^{-\beta}}{\sum RR}.\label{eq:hccc} 
 \end{eqnarray}
If the two parameters,  $A_\omega$ and  $\beta$, are mathematically separable, 
Equation~\ref{eq:hccc} can be rewritten as, 
\begin{eqnarray} 
\omega_\Omega &\approx& A_\omega ~R_\Omega (\beta), \\
R_\Omega (\beta)   &\equiv&  \frac{\sum RR ~\theta^{-\beta}}{\sum RR},\label{eq:rr}
\end{eqnarray}
where we refer to $R_\Omega (\beta)$ as ``random pair function'' (RPF). 
Defying the simple definition of RPF, 
there is a delicate divergence issue, which is described in Appendix \ref{sec:appendixa}. 
The final result on the divergence is that 
RPF, $R_\Omega (\beta)$, is well-defined for $ 0 \le \beta < 1$. 
In this valid $\beta$ range, the coupled equations can be rewritten as,  
\begin{eqnarray}
\omega_{LS}(\theta) &=& \frac{ \theta^{-\beta}  - R_\Omega (\beta) }{A_\omega^{-1}+ R_\Omega (\beta)}. \label{eq:rpf}
\end{eqnarray}
Consequently, the problem of integral constraint is reduced to a self-consistent non-linear fit with the two parameters, 
$( A_\omega^{-1}, \beta )$.

\subsubsection{Best-fit Parameters and Real-space Correlation Lengths}

\begin{figure*}[t]
\centering
\includegraphics[height=3.8 in]{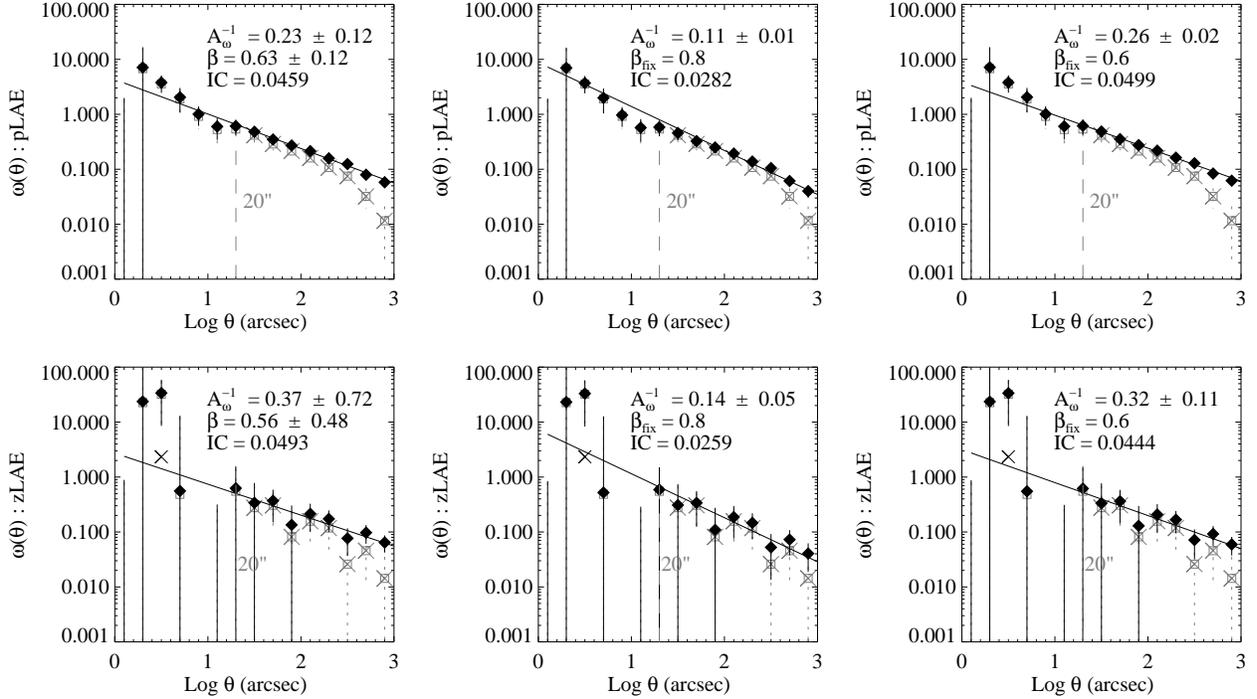}
\caption{The single power-law fits for pLAE (top) and zLAE (bottom) using the non-linear self-consistent fit 
in Equation~\ref{eq:rpf} (left) and the fixed $\beta$ values with $\beta_{fix} = 0.8$ (middle) and $\beta_{fix} = 0.6$ (right). 
The results are summarized in Table~\ref{tableone}.
}\label{fig:5}
\end{figure*}

Figure~\ref{fig:5} shows the results of best-fit parameters for $\theta > 20$\arcsec, 
based on the assumption of single power-law correlation.   
The left panels show the best-fit parameters for pLAE and zLAE, 
using Equation~\ref{eq:rpf} with the RPFs. 
Instead of the conventional fiducial value of $\beta = 0.8$, 
our nonlinear fits predict the slope near $\beta = 0.6$. Hence, we also perform two other fits 
by fixing the slopes, $\beta_{fix} = 0.8$ (middle panels) and $\beta_{fix} = 0.6$ (right panels). 

When the power-law shape of  angular correlation function, $\omega(\theta) = A_\omega \theta^{-\beta}$, is known,  
we can also find its real-space clustering, $\xi (r) =  (r/r_0)^{-\gamma}$, using the Limber equation 
(Peebles 1980; Efstathiou et al. 1991), 
\begin{eqnarray}
\beta  & = & \gamma - 1,\\
A_\omega & = & C r_0 ^{\gamma} \int_{0}^{\infty} F(z) D_\theta^{1-\gamma}(z) N(z)^2 g(z) dz \nonumber \\
& \times &\Big[ \int_0^{\infty} N(z) dz \Big]^{-2}, 
\end{eqnarray}
where $D_\theta(z)$ is the angular diameter distance, 
$F(z)$ the redshift dependence of $\xi (r)$, $N(z)$ the redshift selection function from the zLAE sample, and 
\begin{eqnarray}
g(z) & = & \frac{H_0}{c}\Big[ (1+z)^2 (1+\Omega_M z + \Omega_\Lambda [(1+z)^{-2} - 1])^{\frac{1}{2}} \Big], \\ 
 C & = & \sqrt{\pi} \frac{\Gamma[(\gamma - 1)/2]}{\Gamma(\gamma/2)}. 
\end{eqnarray}  

We summarize the best-fit parameters and related clustering lengths in Table~\ref{tableone}.  
Overall, the parameter ranges of $A_\omega$ and $\beta$ are quite large,  
while the predictions of $r_0$ are relatively consistent as $r_0 \approx 4 h^{-1}$ Mpc. 
The large uncertainties on $A_\omega$ and $\beta$ arise 
from uncertainties in the power-law slope, which in turn are affected by a power-law being 
a poor representation of the observed angular correlation function (cf. the inflection point at 20\arcsec).
As presented in Table~\ref{tableone}, power-law fits with shallower or steeper slopes (i.e., $\beta=0.6,0.8$), 
result in smaller or larger clustering amplitudes (i.e., $A_\omega\approx 4 - 9$, at the consistent result of $\xi (r = 4 h^{-1}$Mpc$) = 1$).

\begin{figure*}[t]
\centering
\includegraphics[height=2.3 in]{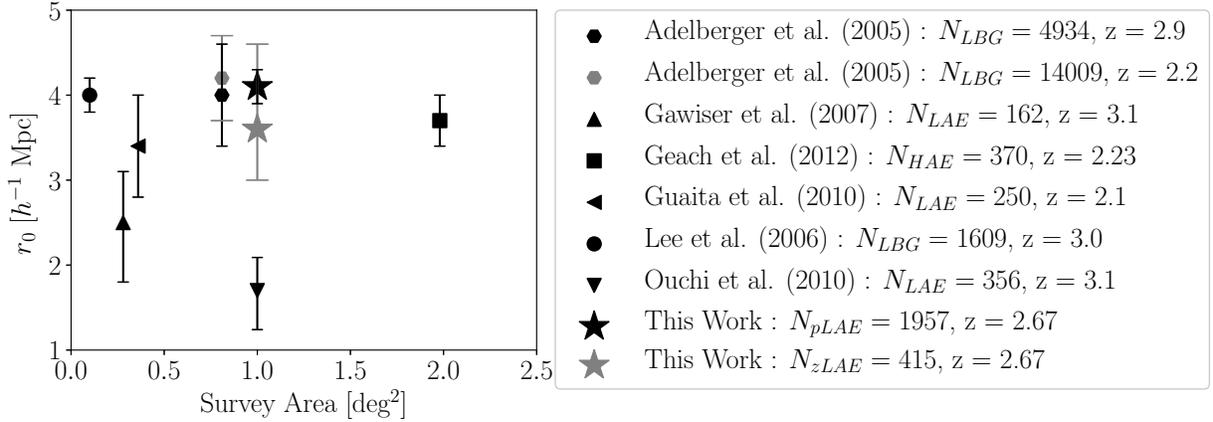}
\caption{The clustering length, $r_0$, vs. survey area for previous studies with similar redshifts. 
Overall, the Bo\"otes LAEs show a comparable or slightly larger clustering amplitude, compared to the previous results. 
}\label{fig:55}
\end{figure*}

The measured clustering length, $r_0 \approx 4 h^{-1}$ Mpc, of the Bo\"otes LAEs at $z = 2.67$, is 
comparable to that derived for LAEs at $z=2.1$ (Guaita et al. 2010), H$\alpha$ emitters at $z=2.23$ (Geach et al. 2012), and the LBGs at $z\approx3$ (Adelberger et al. 2005, Lee et al. 2006), 
and relatively larger than the LAEs from Gawiser et al. (2007) and Ouchi et al. (2010) at $z = 3.1$.  
We summarize these comparisons in Figure~\ref{fig:55}.  
Overall, the Bo\"otes LAEs show a similar or slightly larger clustering amplitude, compared to the previous studies.  

\addtolength{\tabcolsep}{-1pt}  
\begin{deluxetable}{rrrrr}
\tabletypesize{\scriptsize}
\tablecaption{Single Power-law Fits \label{tableone}}
\tablewidth{0pt}
\tablehead{
\colhead{Sample} & \colhead{$A_\omega$} & \colhead{$\beta$} & \colhead{$IC$} & \colhead{$r_0$} \\
\colhead{} & \colhead{(at 1 arcsec)} & \colhead{} & \colhead{} & \colhead{$h^{-1}$ Mpc} 
}
\startdata
pLAE & $4.35^{+4.74}_{-1.49}$ & $0.63 \pm 0.12$ & 0.0459 & $4.1^{+2.3}_{-0.9}$\\
         & $9.09^{+0.91}_{-0.76}$ & 0.8 (fix) & 0.0282 & $4.1^{+0.2}_{-0.2}$ \\
         & $3.85^{+0.32}_{-0.27}$ & 0.6 (fix) & 0.0499 & $4.1^{+0.2}_{-0.2}$ \\
\hline
zLAE & $2.70^{+\infty}_{-1.79}$\tablenotemark{$\dagger$} & $0.56 \pm 0.48$ & 0.0493 & $3.6^{+\infty}_{-1.8}$ \\
         & $7.14^{+3.97}_{-1.88}$ & 0.8 (fix) & 0.0259 & $3.6^{+1.0}_{-0.6}$ \\
         & $3.13^{+1.64}_{-0.78}$ & 0.6 (fix) & 0.0444 & $3.6^{+1.1}_{-0.6}$ \\
\enddata
\tablenotetext{$\dagger$}{Since Equation~\ref{eq:rpf} is only valid for $A_\omega \ge 0$ and $0 \le \beta < 1$, 
the negative range in $A^{-1}_\omega = 0.37 \pm 0.72$ is not mathematically meaningful. 
Hence, we take the upper bound of $A_\omega$ as infinity. } 
\end{deluxetable}
\addtolength{\tabcolsep}{1pt}

\subsection{Interpretations from Mean Halo Occupation Functions}\label{sec:hod}

In the previous section, we have assumed that  the galaxy correlation function 
follows a single power law and measured the amplitude and slope 
by fits to the LAE pair distribution. Historically,  this single power law assumption 
arises from two observations: (1) low redshift galaxies indeed show single power law clusterings in many cases, 
and (2), when a survey volume is small, clustering measurements at small scales are quite uncertain. 

In the current paradigm of hierarchical galaxy formation and evolution, 
observed galaxy clustering (or, galaxy power spectra in $k-$space) 
can be reproduced analytically by using halo occupation distributions 
(HODs; e.g., Seljak 2000, Berlind \& Weinberg 2002, Hamana et al. 2004, Zehavi et al. 2004, Zheng et al. 2005, Lee et al. 2006, Kova{\v c} et al. 2007, Lee et al. 2009, Ouchi et al. 2010, Geach et al. 2012). 
In this HOD formulation, galaxy clustering is generally scale-dependent, deviated from single power-laws, 
due to the non-linear bias in galaxy formation.   

Although this analytic HOD formulation is advantageous to easily reproduce 
observed galaxy clustering analytically, along with intrinsic scale-dependent features, 
it relies on the assumption that the mean halo occupation only depends on the halo mass, 
and is valid when averaged over all halos. 
Effects other than halo mass are generally ignored in the HOD formulation. 
Since we find many $10^{11} M_\odot$ haloes in clusters, filaments, and outskirts around voids,  
it is not likely that all galaxies form in the same way in such different topological environments 
(Hong \& Dey 2015, de Regt et al. 2018). 

\subsubsection{Halo Occupation Function}

In this paper, we adopt the HOD from Geach et al. (2012), 
used for H$\alpha$ emitters  at z=2.23. 
We refer the reader to Appendix B for details regarding this choice. 
The Geach et al. HOD is defined as follows:  
\begin{eqnarray}
N_c(M) &=& F_c^B(1-F_c^A) \exp\Big[ - \frac{\log(M/M_c) ^2}{2 \sigma^2_{\log M}} \Big] \nonumber \\ 
&+& F_c^A\Big[ 1 + \textrm{erf} \Big(\frac{\log(M/M_c)}{ \sigma_{\log M} } \Big) \Big], \label{eq:cen} \\ 
N_s(M) &=& F_s\Big[ 1 + \textrm{erf} \Big(\frac{\log(M/M_{1})}{ \delta_{\log M} } \Big) \Big] \Big( \frac{M}{M_{1}} \Big)^\alpha, \label{eq:sat}\\
N_g(M) &=& N_c(M) + N_s(M),
\end{eqnarray}
where $N_c(M)$ represents the central distribution as a function of halo mass $M$, 
$N_s(M)$ the satellite distribution, 
and $N_g(M)$ the total galaxy counts, for a given halo mass, $M$. 
The central distribution is written using two terms: a Gaussian component centered 
at halo mass $M_c$  with the width of $\sigma_{\log M}$; 
and a smoothed step function component using an Error function with the smoothed length of $\delta_{\log M}$. 
$F_c^A$ and $F_c^B$ represent the duty cycle of central LAEs. 
The satellite distribution is written using the conventional power-law component 
with a tunable satellite's duty cycle, $F_s$. 
Overall, the adopted HOD follows the conventional description of step function centrals 
and power law satellites, with additional flexibility in functional degrees of freedom. 

When considering the complexity of halo occupations for emission line galaxies, 
we need to allow more flexible HODs for LAEs than typical galaxies, selected by 
broad-band photometry, traced by the longer lasting and more consistent emitting source, stars. 
However, over-flexible models inevitably overfit the data; hence, 
they cause degeneracy in possible interpretations. 
In the context of statistical learning, this is an inevitable trade-off 
between flexibility and interpretability of parametric model (James et al. 2013). 
Since we do not have definitive constraints on the HOD for LAEs, 
we will use the HOD from Geach et al. and accept all non-rejected HOD models 
as possible scenarios. 
In Appendix~\ref{sec:appendixpara}, 
we present results from the conventional 3 parameters' HOD (e.g., Zehavi et al. 2005) 
and discuss more about this trade-off issue.

Given the HOD, $N_g(M)$, we derive the galaxy number density, $n_g$, using 
the halo mass function, $n(M)$,  
\begin{eqnarray}
n_g &=& \int N_g(M) n(M) dM.   
\end{eqnarray}
If we have a redshift selection function, $N(z)$, as shown in Figure~\ref{fig:2}, 
then we can take an effective average, $\langle n_g \rangle$, over the selection function as, 
\begin{eqnarray}
\langle n_g \rangle &=& \frac{\int n_g(z) N(z) (dV/dz) dz}{\int N(z) (dV/dz) dz}. \label{eq:den}
\end{eqnarray}
We measure $\langle n_g \rangle = 2.50 \pm 0.05 \times 10^{-4} h^3~\textrm{Mpc}^{-3}$ from 
the $1274 \pm 17$ LAEs, extrapolated using the current yield fraction 65\% from 1957 LAE candidates, 
based on the binomial trials. 
We use the PYTHON package, {\it halomod} (Murray et al. 2013), for HOD calculations 
with the cosmological parameters from {\it Planck13}~ in ASTROPY 
and adopt the halo mass function from Tinker et al. (2008).

\subsubsection{The Best-Fit Parameters : Inverse Correction of Integral Constant}

Our adopted HOD has 8 parameters, $M_c, M_1, \alpha, F^A_c, F^B_c, F_s, \sigma_{\log M}$, and $\delta_{\log M}$. 
Generally, the smoothing scales of step function terms, $ \sigma_{\log M}$, and $\delta_{\log M}$, 
are not as critical to the shape of the resulting angular correlation function as $\alpha, M_1$, and $M_c$.
This means that $\alpha, M_1$, and $M_c$ are well constrained by the angular correlation measurement, 
whereas the others are not. 
Bayesian samplings can provide quantitative information, 
about how well the observed angular correlation function can constrain each parameter, 
by studying the posterior probability density function (PDF) as shown in Figure~\ref{fig:mcmc}.

Among the 8 parameters, we first fix  one of the least important parameters, $\delta_{\log M} \equiv 1$, 
which controls the width of error function in Equation~\ref{eq:sat}.
Since $\sigma_{\log M}$ is used in both Gaussian and Error functions in Equation~\ref{eq:cen}, 
we do not fix this parameter to allow the Gaussian width to vary. 
From the density normalization of Equation~\ref{eq:den}, $M_c$ can be determined using 
$\langle n_g \rangle = 0.0025 ~ h^3~\textrm{Mpc}^{-3}$ . 
Therefore, our final HOD has the 6 free parameters, $M_1, \alpha, F^A_c, F^B_c, F_s, \sigma_{\log M}$.

As we have pointed out in \S \ref{sec:w}, IC can be determined only by its true angular correlation function. 
For single power-law correlations, we can resolve this IC problem using the non-linear fit with 
random pair function in Equation~\ref{eq:rpf}. 

In the HOD formulation, we can resolve this issue using the inverse correction of integral constraint as follows. 
First, we have a well-defined model prediction of the angular correlation function, $\omega_{HOD}(\theta)$, from a given HOD.  
Since this is a true angular correlation function, not degraded by survey volume, 
we can calculate its corresponding IC, $\omega_\Omega$, directly from $\omega_{HOD}(\theta)$ : 
\begin{eqnarray}
\omega_\Omega &\approx& \frac{\sum RR ~ \omega_{HOD}(\theta)}{\sum RR}. 
\end{eqnarray}
From $\omega_\Omega$, we define a new inverse HOD angular correlation function, $\tilde{\omega}_{HOD}(\theta)$,  as :
\begin{eqnarray}
\tilde{\omega}_{HOD}(\theta) &\equiv& \frac{  \omega_{HOD}(\theta) - \omega_\Omega}{ 1 + \omega_\Omega}. \label{eq:iic}
\end{eqnarray}
This {\it inversely} corrected HOD function, $\tilde{\omega}_{HOD}(\theta)$, is now 
directly comparable to the observed LS estimator, $\omega_{LS}(\theta)$. 
Therefore, we can write down the correct $\chi^2$ as 
\begin{eqnarray}
\chi^{2}(M_1, \alpha, F^A_c, F^B_c, F_s, \sigma_{\log M}) & &  \nonumber \\ 
= \sum_{i} \frac{  \big[\omega_{LS}(\theta_i) - \tilde{\omega}_{HOD}(\theta_i) \big]^2}{ \sigma^2_{LS}(\theta_i)} & &,
\end{eqnarray}
where $\theta_i$ represents each angular bin and $\sigma^2_{LS}$ the bootstrap sampling variance of LS estimator. 
Finally, we define the likelihood function as 
\begin{eqnarray}
\ln \mathcal{L} & = & - \frac{1}{2} \chi^{2}. 
\end{eqnarray}



\subsubsection{Results : Degeneracy in Two--point Statistics}\label{sec:degenerate}

We use two different methods to obtain best-fit HOD parameters, 
(1) one from $\chi^2$ minimization, referred to as Model\#1, and 
(2) the other from Bayesian posterior probability density function (PDF), referred to as Model\#2,  
obtained using the Markov chain Monte Carlo (MCMC) sampler, {\it emcee} 
(Foreman-Mackey et al. 2013). 

\begin{figure*}[t]
\centering
\includegraphics[height=6.7 in]{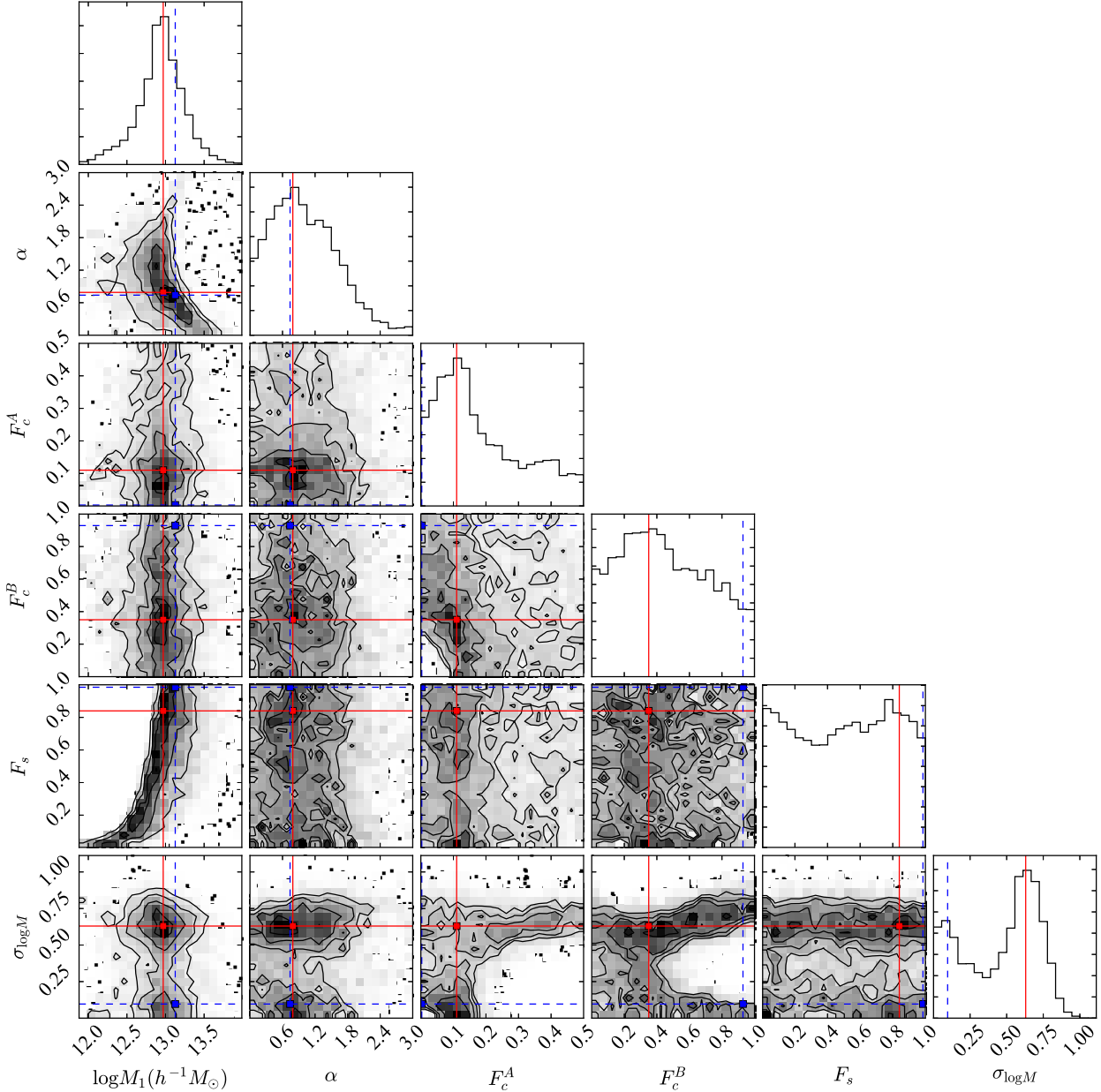}
\caption{The posterior probability density function from the MCMC run. 
This 6 dimensional posterior probability density function is visualized in contours (2D marginalized probabilities) 
and histograms (1D marginalized probabilities). 
The blue points and dashed lines represent the parameters selected from the least chi-square method (Model\#1),  
and the red points and solid lines selected from the posterior probability function (Model\#2). 
The parameters are summarized in Table~\ref{tabletwo}. 
}\label{fig:mcmc}
\end{figure*}

Figure~\ref{fig:mcmc} shows the result of posterior PDF, 
obtained using the MCMC sampler, {\it emcee}. 
We put 120 walkers in total (i.e., 20 walkers for each parameter) and iterate 950 steps. 
We discard the early 450 steps as burn-in and take 500 steps to retrieve the posterior PDF.  
This 6 dimensional posterior PDF is visualized in contours (2D marginalized probabilities) 
and histograms (1D marginalized probabilities). 
The median value and $\pm 1 \sigma$ errors for each parameter  
from the 1D marginalized histograms are listed in Table~\ref{tabletwo}. 
Since the marginalized distributions for $F_s$ and $\sigma_{\log M}$ are flat and bimodal respectively, 
it is not informative to present the medians and errors for these parameters. 
To interpret the posterior PDF, $\log_{10} M_1$ is the only parameter, well constrained by the angular correlation function. 
The others are marginally (or poorly) constrained. 
This is not a surprising result when considering the relatively large number of free parameters  
compared to the conventional 3 parameters' HOD. Though there are many other statistics 
such as Minkowski functionals, genus, percolation threshold and higher order correlation functions, 
the current HOD formulation only fits the abundance and two-point statistic of observed populations.  
Hence, the degeneracy in HOD models is inevitable if the number of free parameters exceeds the constraining power 
of abundance and two-point correlation; i.e., if the HOD function is over-flexible. 

The posterior PDF provides a better statistical interpretation for the best fit model than other methods 
such as maximum likelihood or least chi-square. 
However, since the least chi-square method is widely used, we also compute it; 
hence, Model\#1,  using the Nelder-Mead method implemented in the PYTHON/SCIPY package. 
From various initial positions, we obtain the consistent output of Model\#1. 
However, we cannot reject the possibility that Model\#1 is derived from a local minimum. 
We take, therefore, Model\#1 as one of many possible selections, 
statistically allowed within the posterior PDF. 
The blue points and dotted lines in Figure~\ref{fig:mcmc} represent the location of Model\#1 in the parameter space. 
Though this location is less likely in the posterior PDF, this location in parameter space 
is not ruled out by the MCMC approach.

From the 2D contours in Figure~\ref{fig:mcmc}, we select a more likely position, Model \#2,  represented 
by the red points and solid lines. 
A major difference between Model\#1 and Model\#2 comes from the parameter, $\sigma_{\log M}$ , 
which shows the bimodal histogram. Model\#1 is selected 
from the minor bump, while Model\#2 from the major bump. 
These Bayesian selections contrast 
with the different reduced chi-square values, $\chi^2 / \nu  = 0.51$ for Model\#1 
and $\chi^2 / \nu  = 1.2$ for Model\#2. 
Model\#1 is a preferred choice, therefore, in the least chi-square method, 
whereas Model\#2 in the Bayesian method. 
The is issue is that neither model is rejected by the tests
in abundance and two-point statistic, 
though their HODs are significantly different. 
 
\begin{figure*}[t]
\centering
\includegraphics[height=2.5 in]{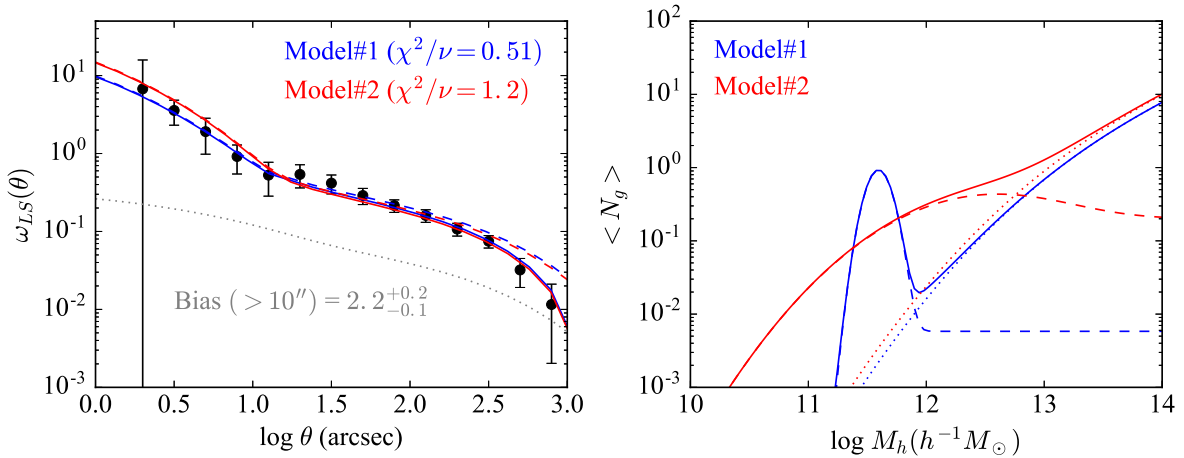}
\caption{ The two point correlation functions (left) and halo occupation distributions (HODs; right) for Model\#1 (blue) and  Model\#2 (red). 
The dotted grey line represents the angular dark matter correlation function (Takahashi et al. 2012). 
On the left panel, each dashed line represents the true two point function from each HOD Model without any effect of survey volume size  
and each solid line the inversely corrected two point function using our inverse integral constraint method. 
In the right panel, the dashed, dotted and solid lines represent the expected number 
distributions of central galaxies, satellite galaxies and the total, respectively.
The two models, Model\#1 and Model\#2, show very different HODs, but both predict similar two-point correlation functions, matching 
the observed clustering in the accuracy of practical studies; or, at least, both are not statistically rejected in the test of two-point statistics. 
}\label{fig:hod}
\end{figure*}

Figure~\ref{fig:hod} shows the angular two point correlation functions (left) and halo occupation distributions 
(HODs; right) for Model \#1 (blue) and Model \#2 (red). 
The dotted grey line represents the angular dark matter correlation function (Takahashi et al. 2012) 
and we measure the bias, $b_{LAE} = 2.2^{+0.2}_{-0.1}$, at scales larger than 10\arcsec. 
This is slightly larger than, but still consistent with that found by previous works 
($b\approx 1.5-2.0$; Gawiser et al. 2007, Guaita et al. 2010, Ouchi et al. 2010, Lee et al. 2014).
In the left panel, each dashed line represents the two point function from each HOD Model 
without the effect of survey volume size and each solid line the inversely corrected two point function 
using our inverse integral constraint method. 
After this inverse correction, the observed clustering points from the LS estimator 
(black solid circles with error bars) can be 
directly comparable to the solid line; i.e., we can use the LS estimator as a direct observable 
without any further correction. 
In the right panel, the dashed, dotted and solid lines represent the expected number 
of central, satellite and total galaxies, respectively.

In this figure, Model\#1 and Model\#2 show very different HODs, 
especially in the central galaxy populations. 
For Model\#1, the central LAEs are mostly occupied in a very narrow halo mass range, 
centered at $\log_{10} M_c = 11.6$ with the Gaussian width of $\sigma_{\log M} = 0.097$. 
At its peak, the occupation fraction, $< N_g >$, reaches 93\%. 
This drops rapidly as the halo mass increases or decreases away from this peak halo mass. 
For massive halos over $10^{12} h^{-1} M_\odot$, the central LAE occupations become less than 0.3\%. 
Therefore, most of LAEs found in these massive halos should be satellites in this Model\#1 scenario; 
hence, we refer to this as, namely, ``Dead Core Scenario" or ``Dusty Core Scenario". 
The lack of central LAEs at halo masses $>~10^{12-13}$ in Model\#1 may imply 
that central galaxies in these halos do not produce much Ly$\alpha$ emission, 
either because they are more rapidly quenched or that they are dustier on average; hence, dead or dusty cores. 

 On the other hand, for Model\#2, the central LAEs are distributed over a broad range of halo masses, 
centered at $\log_{10} M_c = 12.40$ with the Gaussian width of $\sigma_{\log M} = 0.63$. 
At this Gaussian peak, the occupation fraction is 31\%, which is much lower than 
the dominant 93\% from the Dead Core Scenario. 
For massive haloes, even larger than $10^{13} h^{-1} M_\odot$, 
the central occupation fractions are above 20\% in Model\#2. 
We refer to this as ``Active Core Scenario" or ``Pristine  Core Scenario'', suggesting 
that the central galaxies in massive halos are still less contaminated by dust, 
actively emitting Ly$\alpha$ photons, unlike the dead or dusty cores from Model\#1. 

Consequently, Model\#1 and Model\#2 suggest very different scenarios 
about the formation and evolution of LAEs at $z \approx 2.67$. 
We cannot discern which scenario is more reliable for the observed LAEs at $z \approx 2.67$. 
To resolve this issue, we need to resort to higher order correlations, 
which in turn are limited by the sample statistics.

\begin{deluxetable*}{crrrrrr}
\tabletypesize{\scriptsize}
\tablecaption{The Parameters of Halo Occupation Functions \label{tabletwo}}
\tablewidth{0pt}
\tablehead{
\colhead{Name} & \colhead{$\log_{10} M_1$} & \colhead{$\alpha$} & \colhead{$F^A_c$} & \colhead{$F^B_c$} & \colhead{$F_s$}  & \colhead{$\sigma_{\log M}$}
}
\startdata
Model\#1\tablenotemark{a} & 13.13 & 0.74 & $2.9\times 10^{-3}$ & 0.93 & 0.99 & $9.7\times 10^{-2}$\\
Model\#2\tablenotemark{b} & 12.97 & 0.79 & 0.11 & 0.35 & 0.84 & 0.63\\
\hline
 & &  &  &  &  & \\
Posterior PDF\tablenotemark{$\dagger$} & $12.95^{+0.26}_{-0.29}$ & $0.94^{+0.69}_{-0.56}$ & $0.14^{+0.21}_{-0.09}$ & $0.43^{+0.34}_{-0.27}$ & flat & bimodal\\
\enddata
\tablenotetext{a}{From the density normalization,  $\log_{10} M_c = 11.59$ for Model\#1.}
\tablenotetext{b}{From the density normalization,  $\log_{10} M_c = 12.40$ for Model\#2.}
\tablenotetext{$\dagger$}{We present the median value for each parameter with $\pm 1 \sigma$ errors 
from the posterior PDF, shown in Figure~\ref{fig:mcmc}. Since the marginalized distributions for $F_s$ and $\sigma_{\log M}$ 
are flat and bimodal respectively, it is not informative to present the medians and errors for these parameters. } 
\end{deluxetable*}

\section{Statistics of Network Topology}\label{sec:network}

In the previous section we have presented measurements of the two-point correlation function 
and abundance of LAEs. The measurements are fit by two HOD models 
which predict the same abundance and two-point correlation within the uncertainties. 
However, their HODs are very different, 
especially in the central galaxy populations. 
This is an evident degeneracy in two-point statistics. 

In this section, we use network science tools to investigate the topological structures 
of the observed LAEs (Observed LAEs), and compare these with the topologies generated 
by the two best-fit HOD models (Model\#1 and Model\#2) and random spatial distributions (Random Model).
From the statistics of network topology, we show that both Model\#1 and Model\#2 fail to explain 
the spatial distribution of observed LAEs; hence, the topological structures of observed LAEs 
are different from the HOD models' predictions. 
This indicates that the assumption of constant halo occupation for all halos of a given mass is too simple 
to be applicable, at least, to LAEs.

\subsection{Generating Networks from Galaxy Point Distributions}

We generate 60 mocks for each HOD model by populating LAEs using the halo catalog 
from Small MultiDark Planck simulation (SMDPL; Klypin et al. 2014) and projecting them on the sky mask, shown in Figure~\ref{fig:1}. 
Central galaxies are randomly placed in parent haloes given by the HOD. 
Likewise, satellite galaxies are placed in their sub-haloes to match the target occupation. 
Note that our catalog allows for satellites to be placed in parent haloes that may or may not host a central galaxy.

A single mock catalog matches the area of the survey. The depth is given by the $IA445$ filter transmission curve, 
which defines a redshift and comoving distance range where the Ly$\alpha$ line falls inside the filter. 
Then, multiple mock catalogs are extracted from the simulation volume with no overlapping. 
The different number of galaxies and clustering in each of the mocks is thus a result of cosmic variance. 

For observed LAEs, we have measured the redshifts of 635 candidates from the total 1957 pLAEs. 
For the rest of 1322 photometric candidates, since the  current yield fraction is 65\%, 
we generate a binomial ensemble with 300 realizations for considering the incompleteness 
of our  spectroscopic followup. This ensemble size is large enough to show asymptotic behaviors 
in graph statistics; i.e., no quantitative differences in graph statistics by taking larger ensemble sizes.   
In \S \ref{sec:netresult}, we will present the details about this binomial convergence. 
Finally, as a basic comparison set, we generate 60 random point distributions as Random Model.


From each spatial distribution, we build a network using the conventional Friends-of-Friends (FOF) recipe 
(Huchra \& Geller 1982, Hong \& Dey 2015, Hong et al. 2016) for a given linking length $l$, where the {\it adjacency matrix} is defined as, 
\begin{equation}\label{eq:adj}
A_{ij} = \left\{ \begin{array}{ll}
	1 & \textrm{   if   } r_{ij} \le l,  \\
	0 & \textrm{   otherwise, } 
	\end{array} \right.
\end{equation}
where $r_{ij}$ is the distance between the two vertices (i.e., galaxies), $i$ and $j$. 
This binary matrix quantitatively represents the network connectivities of the FOF recipe. 
Many important network measures are derived from this matrix.
Interested readers are directed to Newman (2003), Dorogovtsev, Goltsev \& Mendes (2008), and  
Barth\'{e}lemy (2011) for further information.

Figure~\ref{fig:10b} shows the histogram of the number of galaxies, $N_{galaxy}$, 
for each model composed of 60 mocks; 
Observed LAEs (red), Random Model (grey), Model\#1 (green)  and  Model\#2 (blue). 
For a proper comparison, we take 60 binomial samples from the total 300 realizations for this histogram; 
though no qualitative difference in abundance statistics between 60 and 300 binomial realizations. 
The variances of $N_{galaxy}$ for Model\#1 and Model\#1 are due to cosmic density fluctuations, 
confined by the size of survey volume. 
Observed LAEs are shown at a range of possible abundances estimated 
using the known photometric  uncertainties and spectroscopic completeness, 
which suggest an LAE abundance in the field of 1274$\pm$17. 
Finally, the variance for Random Model is Poissonian, a comparable random reference   
to the other models. 

The cosmic variances of Model\#1 and Model\#2 are much larger 
than the binomial variances of Observed LAEs and Random Model. 
We, therefore, expect the network properties of the observed LAEs 
to be contained within the range exhibited by the HOD mocks.

\subsection{Results : Implications from Network Statistics}\label{sec:netresult}

\begin{figure}[t]
\centering
\includegraphics[height=2.0 in]{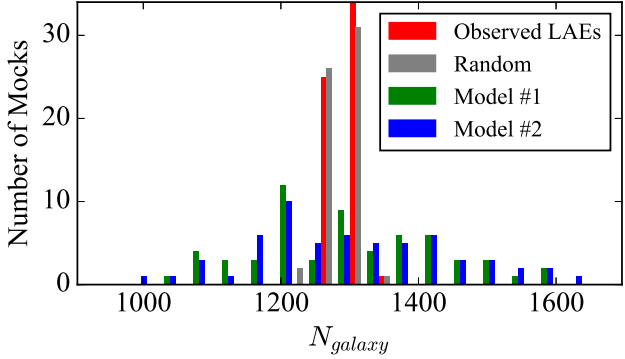}
\caption{The histogram of the number of galaxies, $N_{galaxy}$, for each model, composed of 60 mocks; 
Observed LAEs (red), Random Model (grey), Model\#1 (green)  and  Model\#2 (blue). 
The cosmic variances of Model\#1 and Model\#2 are much larger than 
the binomial variances of Observed LAEs and Random Model. 
Naively, due to this dominance of cosmic variance, 
we may expect that the network statistics from Observed LAEs are fully 
embedded within the cosmic variances of the HOD mocks. 
}\label{fig:10b}
\end{figure}  

\begin{figure}[t]
\centering
\includegraphics[height=2.5 in]{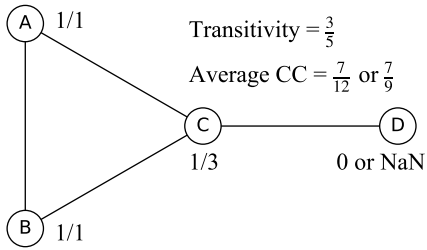}
\caption{  The graph schema demonstrating the meanings of transitivity and average clustering coefficient (Average CC). 
In this graph, we can find 5 triplets (i.e., 5 $\vee$ configurations), 3 from the `ABC' triangle 
and the other 2 from the ``Y" shape, connected from the vertex `D' with the pivot center `C'.   
Among these 5 triplets, three of them on the `ABC' triangle are closed. Hence, the transitivity of this graph is $\frac{3}{5}$. 
The clustering coefficient is a transitivity-like quantity, but assigned to each vertex. 
For example, the vertex `C' has three neighbors, `A', `B', and `D'; hence, 3 $\vee$ configurations, 
centered on `C'. In general, for a vertex with $k$ neighbors, $\frac{k(k-1)}{2}$ triplet combinations exist. 
Since only `A' and `B' vertices are connected among the three triplets, the clustering coefficient for `C' 
is $\frac{1}{3}$. Similarly, 1 is the clustering coefficient for each of `A' and `B'. 
For a vertex with $k < 2$ neighbors, we cannot define a clustering coefficient since the denominator is zero. 
In this case, to the vertex, we can assign (1) 0 or (2) not-a-number (NaN). For the former, the vertex `D' is counted 
when averaging all clustering coefficients, resulting in the Average CC $= \frac{7}{12}$, 
while, for the latter, `D' is excluded, the Average CC $= \frac{7}{9}$. 
In this paper, we choose the latter definition to assign NaNs to all vertices with $k  < 2$ neighbors.
}\label{fig:10c}
\end{figure}

\begin{figure*}[t]
\centering
\includegraphics[height=7.5 in]{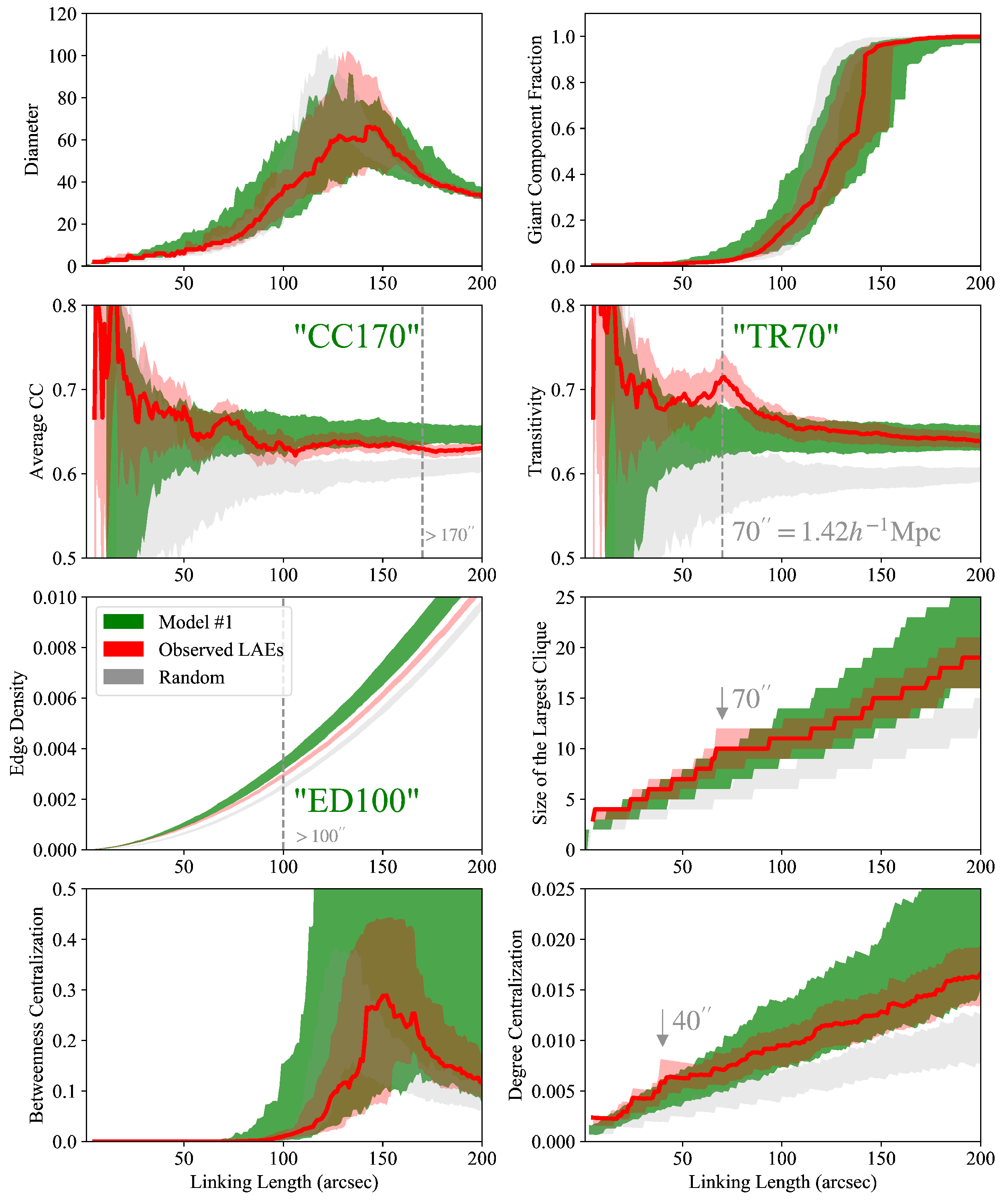}
\caption{The network measurements for Model\#1 (green), observed LAEs (red), 
and random point distributions (grey); diameter (top-left), giant component fraction (top-middle), 
transitivity (top-right), average clustering coefficient (middle-left), 
edge density (center), size of the largest clique (middle-right), 
betweenness centralization (bottom-left), closeness centralization (bottom-middle), 
and degree centralization (bottom-right). 
See the text for details. 
}\label{fig:11}
\end{figure*}  

\begin{figure*}[t]
\centering
\includegraphics[height=7.5 in]{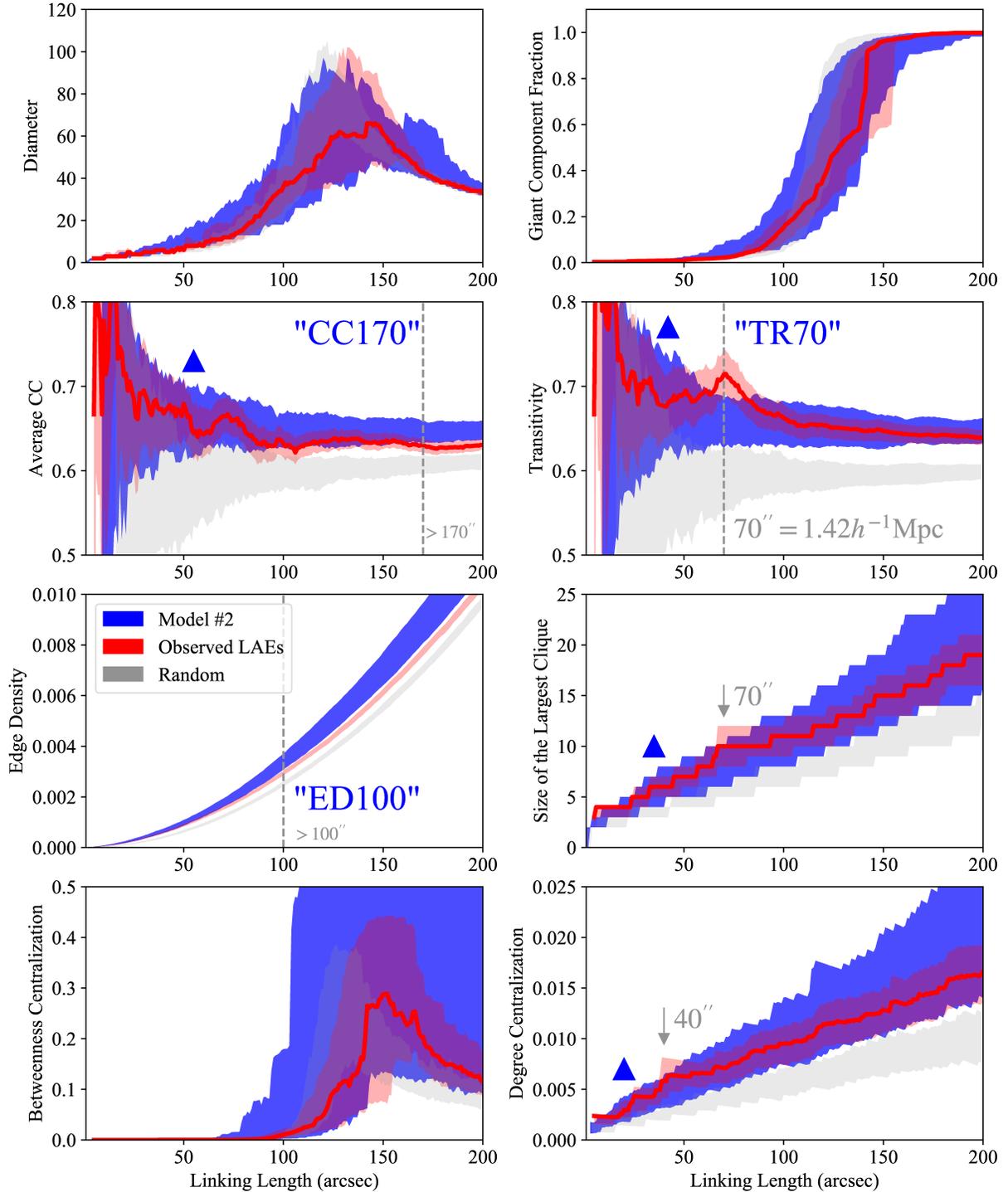}
\caption{The same with Figure~\ref{fig:11} for Model\#2 (blue), observed LAEs (red), and random point distributions (grey). 
The 4 blue solid triangles indicate the regions of improved statistics by Model\#2. 
The major difference between Model\#2 and Model\#1 is the higher fraction 
of central galaxy occupation in massive halos (see Figure~\ref{fig:hod}).
}\label{fig:12}
\end{figure*}  

\begin{figure*}[t]
\centering
\includegraphics[height=3.0 in]{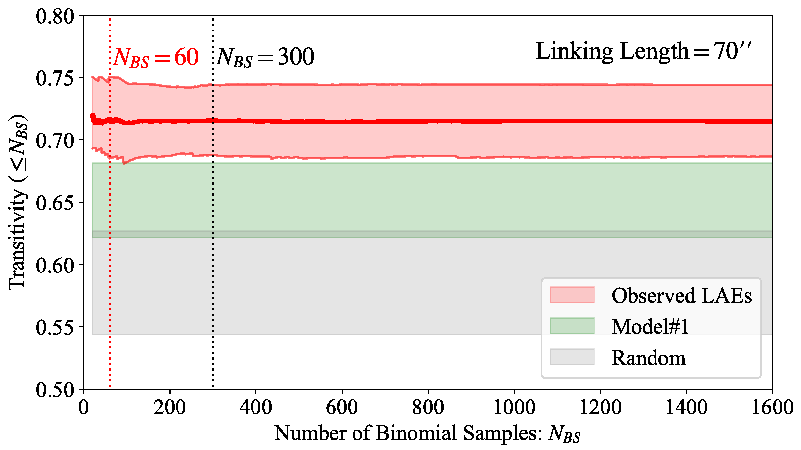}
\caption{The transitivity measurements vs. the number of binomial samples, $N_{BS}$, 
at the linking length of $70\arcsec$, for Observed LAEs (red shaded area); 
hence, a test of the convergence of binomial sampling about the anomaly ``TR70''.  
The central thick line represents the 50 percentile (median)  
and the others 5 and 95 percentiles. 
The red shaded range at $N_{BS} = 300$ (dotted black vertical line) is what is shown at $70\arcsec$ in Figure~\ref{fig:11}. 
For comparison, we also plot the ranges of transitivities for Model\#1 (green) and Random Model (grey) at $70\arcsec$ 
with the same colors shown in Figure~\ref{fig:11}. 
For $N_{BS} \ge 300$, the transitivity measurements show saturated asymptotic behaviors. 
Even for $N_{BS} = 60$, no qualitative differences can be found from the larger sampling sizes for $N_{BS} \ge 300$.
}\label{fig:12b}
\end{figure*}

For various angular linking lengths from 0\arcsec~ to 200 \arcsec, 
we build a series of FOF networks for each spatial distribution. 
Then, for each network, we measure 8 network quantities: 
{\it diameter, giant component fraction, average clustering coefficient} (average CC), 
{\it transitivity, edge density, size of the largest clique, betweenness centralization,} and {\it degree centralization}. 
We present the definitions of these 8 quantities in a separate section, Appendix~\ref{sec:appendixb}, 
so as not to distract the reader from the main thread of this paper.

Figures~\ref{fig:11} and \ref{fig:12} show the results of network statistics 
for the Observed LAEs, Model\#1, Model\#2, and Random Model, as a function of the linking lengths. 
Among 60 realizations for each model (300 realizations for Observed LAEs), 
we remove 5\% outliers on both top and bottom regions. 
Therefore, each colored region represents  90\% of its statistical distribution. 
For Observed LAEs, as a likely position for the case of complete spectroscopic selection, 
we plot the median position, using the red solid line.

To test the reliability of binomial sampling for handling the incomplete spectroscopic survey, 
we measure transitivity values for various numbers of binomial sampling. 
Figure~\ref{fig:12b} shows the transitivities vs. the number of binomial samples, $N_{BS}$, 
at the linking length of $70\arcsec$ for Observed LAEs (red shaded area). 
Notably, the transitivities show asymptotic behaviors for $N_{BS} \ge 300$; 
hence, no further variations for larger sampling sizes. 
Even for $N_{BS} = 60$, there is no qualitative difference from the case of $N_{BS} = 300$. 
We think that this is because the randomness of binomial sampling affects the graph statistics severely and directly.  
As shown in Figure~\ref{fig:10b}, the variance of abundances for Observed LAEs is quite smaller than the HOD mocks. 
However, the variances of graph statistics for Observed LAEs are not much different from the HOD mocks 
even for tens of binomial realizations as shown in Figure~\ref{fig:12b}. 
Hence, though there are $2^{1322}$ kinds of binary permutations (detected or non-detected LAEs) 
for unexplored photometric candidates, the random selections by binomial sampling shuffle the outputs 
quite enough to show the asymptotic statistical behaviors in graph measurements for $N_{BS} \ge 300$. 
We note that this argument is only valid when the best guess of complete spectroscopic survey 
is to extrapolate the current yield to the rest of unexplored photometric candidates. 
We assume that this extrapolation is a practically reasonable approach with the currently available pieces of limited information.

From the results of  network statistics, we obtain the 4 main implications below. 

\subsubsection{Both HOD models fail to explain the graph topology of observed LAEs}

In Figure~\ref{fig:11}, we find that the comparison of network measures computed from the observed data 
with those computed from the mocks results in the following three main differences : 
(a) the transitivity curve of the observed data shows a ``feature'' at a scale of $\approx$70\arcsec\ 
(1.4 $h^{-1}$~Mpc comoving) which is not present in the mocks, and which is not observed 
in the average CC (hereafter, we refer to this anomalous feature as TR70); 
(b) the average CC curve of the observed data at scales $>$170\arcsec\ (3.4 $h^{-1}$~Mpc comoving) 
is lower than that computed for the mocks (hereafter, CC170); and 
(c) the observed  edge density curve at scales $>$100\arcsec\ is not reproduced 
by the mocks (hereafter, ED100). 
Roughly, average CC and transitivity are biased and unbiased triangle densities respectively. 
Figure~\ref{fig:10c} shows a schema demonstrating the meanings of these triangular statistics. 
Edge density is a connection (or, friendship) density, dividing the number of edges 
by the total number of pair-wise combinations.
We discuss each of these anomalies in more detail below.

TR70 is the most conspicuous anomaly in the 8 panels of Figure~\ref{fig:11}. 
Near 70\arcsec, the transitivity of observed LAEs is much higher than the predictions of HOD mocks. 
The boundaries of the shaded regions shown for the models and observed LAEs 
represent the 5\% outliers. Hence, there is $< 0.25$\%  (5\% $\times$ 5\%) chance that this observed feature 
can be reproduced by Model\#1; or over $99.75$\% chance to reject Model\#1 . 
In addition, this TR70 feature is not likely to be a result of the image mask, 
since it is not seen in the transitivity curves constructed from the mocks, to which the same mask is applied.

The angular scale of 70\arcsec ~corresponding to 1.4$h^{-1}$~Mpc in the comoving scale  
is smaller than the typical scales for proto-clusters (Chiang et al. 2013, Orsi et a. 2016), 
but still larger than most of single halo scales. 
Hence, the transitivity excess at this intermediate scale 
suggests a {\it strong intergalactic interaction} in the formation of LAEs in this field.  
We explore this strong environmental effect in more details in a separate section 
with the additional network statistics of {\it clique} and {\it centralization}.

For linking lengths greater than 170\arcsec, the average CCs of 
Observed LAEs are lower than the Model\#1's predictions. 
The average CC is biased to the majority's CC value, 
while the transitivity is a network-wise unbiased triangle density. 
In our Bo\"otes LAEs, field LAEs are the majority, since group LAEs are rare. 
Small neighbors of field LAEs, hence, dominate the average CC statistic. 

Unlike the feature TR70 seen in the transitivity curve, the average CC measurement 
does not show a significant anomaly at a scale of 70\arcsec. 
This suggests that the TR70 anomaly is not likely to be caused by the majority 
of field LAEs but instead by the LAEs in group environments, 
which are a minority of the observed population.
Near 70\arcsec, therefore, the HOD mocks seem to reproduce the 
triangular configurations for {\it field} LAEs, the {\it majority}, 
but fail when including the {\it minority}, group LAEs. 
In other words, something interesting happens in group LAEs near 70\arcsec, 
which cannot be reproduced by the HOD formulation. 

In contrast, the transitivity measure is consistent with the HOD prediction 
at scales $>170$\arcsec, whereas the average CC is not.
This indicates that the observed LAEs and HOD mocks are consistent 
in the network-wise triangle densities at $> 170$\arcsec, 
but the HOD mocks overpredict the average CC values at these scales. 
Namely, the observed {\it field} LAEs are less triangular than the HOD mocks 
in the local clustering configurations at $> 170$\arcsec. 
It is not straightforward to determine which topological configuration causes this feature. 
One possible interpretation is that the observed field LAEs have more obtuse angles 
in triple configurations (i.e., $\vee$) than the HOD mocks. 
These more obtuse configurations can decrease the local CCs. 
As a trade-off, the observed LAEs in group environments 
need to have more triangular configurations, since the transitivity still needs 
to be consistent with the HOD mocks at $> 170$\arcsec. 
Hence, our possible interpretation of CC170 is that the real observed LAEs 
are less triangular with more obtuse angles in spatial alignments of the field environments  
but more triangular in the group environments than the HOD mocks; 
{\it more strained and stretched in field LAEs and more balled and compact in group LAEs than the HOD mocks.}

For the edge density measurements, the HOD mocks overpredict the number of edges at most scales. 
Along with TR70, this is an additional evidence that the HOD mocks fail 
to reproduce the topology of observed LAEs. 
The difference in edge densities is more visible for $>100$\arcsec. Hence, we refer to this anomaly 
as ED100. Since edge is a basic structure, many factors can affect this count of connections. 
The less triangular configuration in field LAEs, mentioned above for interpreting CC170, 
can be one of such factors to lower the edge density than the HOD mocks.

In Figure~\ref{fig:12}, Model\#2 seems to match the network statistics better than Model\#1, 
but the three major anomalies are still not resolved by Model\#2. 
Therefore, both HOD models fail to explain the real graph topology of observed LAEs. 
When considering the simplicity of mean halo theory,  
the HOD mocks explain relatively well the overall topological features of observed LAEs, 
only failing at certain scales. In contrast, the random point distribution fails at most scales in most statistics.

Overall, the anomalies found in network statistics 
suggest that the HOD mocks fail in the topological tests of network statistics; 
or, if the HOD formulation is right, the Bo\"otes LAEs are a very special outlier in the cosmic variance, 
showing very abnormal environmental effects. 
We note that we have only shown the failures of two specific HOD models in graph statistics. 
This could be a suggestive evidence that the current HOD formulation needs to be improved 
for explaining (especially) the populations depending on environments strongly, but 
not a definitive evidence to deny the whole HOD framework.

\subsubsection{The Bo\"otes LAEs are not a good filament/wall tracer}

 In Figures~\ref{fig:11} and~\ref{fig:12}, the 8 panels can be divided into two groups; 
(1) {\it diameter, giant component fraction, betweenness centralization,} 
and (2) {\it average CC, transitivity, edge density, size of the largest clique}. 
The network measures in the first group show no significant statistical differences 
between the observed LAE sample and the other models. 
In contrast, the second group of measures do show differences, as described in the previous section.
We note that the first group reflects {\it the global pathway structures} 
while the second group  {\it the local configurations} as their definitions indicate, 
described in Appendix~\ref{sec:appendixb}.


The observed LAEs and HOD mocks are different in the local topology from random networks,  
while, in the global topology, the HOD mocks seems to even overwhelm the random networks in variance. 
The latter point seems confusing;  
especially, considering the results of our previous study (Hong et al. 2016), which demonstrates 
that simulated galaxies and L\'evy flights show very different topology not only {\it locally} but also {\it globally}. 

This may be due to the transient property of LAEs, having a specific duty cycle. 
For the work of Hong et al. (2016), we selected all simulated galaxies 
with stellar masses greater than $10^8$ M$_\odot$; hence, 
more likely to trace underlying filamentary structures than transient LAEs. 
The HOD recipe of probabilistic occupations on dark matter halos also can
add more stochastic fluctuation to the mock LAEs. 
Analyzing the two-dimensional projection of the large scale distribution 
also dilutes and distorts the signal (the data analyzed in Hong et al. 2016 used the full 3-d distribution).

Consequently, though the observed and mock LAEs show many distinct local features, 
the global large-scale structures such as filaments or walls are not well characterized 
in the 2-dimensional projection of the LAE distribution and will require the complete redshift 
distribution for proper analyses.

\subsubsection{Strong environmental effect on the formation and evolution of LAEs}

In this section, we investigate which topological configuration may be responsible for TR70. 
There are many graph structures, which can increase transitivity. 
One of them is a {\it clique}. 
As explained in Appendix~\ref{sec:appendixb} and shown in Figure~\ref{fig:9}, 
a clique is a complete subgraph, 
and galaxy groups and clusters form cliques in galaxy FOF networks. 
Therefore the abnormal excess in triangular configurations, TR70, 
can be related to clique statistics. 

To test this idea, we measure 
{\it the size of the largest clique}\footnote{Many network algorithms related to cliques 
need long computation times, and some of them are NP-complete. 
Hence, in this paper, we measure one of the basic clique measurements, 
{\it the size of the largest clique} (a.k.a., {\it clique number}), for which some efficient algorithms are known.}, 
shown in the right panels of the third row in Figures~\ref{fig:11} and~\ref{fig:12}.
The excess of the largest clique size is also found at 70\arcsec,  
though its statistical significance is not as strong as TR70. 
The second, third, and next largest cliques also contribute to the transitivity, 
though they are not traced by this measurement. 
Hence, this suggests that TR70 is due to the larger clique sizes in the observed LAEs than in the HOD mocks. 
The median of the HOD mocks predicts that 7 LAE should inhabit the largest clique; 
the observed distribution shows 10, suggesting that scale sizes of 1.4$h^{-1}$~Mpc contain 
$\sim$43\% more LAEs than predicted by the mocks.
TR70 may therefore indicates a strong environmental effect on the formation and evolution of LAEs at $z \approx 2.67$, 
exerted within the scale of $1.4 h^{-1}$ comoving Mpc (at least for the LAEs within this dataset).

If we find a clique excess at a certain scale, 
we can also expect some related feature in {\it centralization} measurements.   
Figure~\ref{fig:10} in Appendix~\ref{sec:appendixb} shows the three graph schemata, {\it ring, star,} and {\it clique}, with 7 vertices. 
These schemata demonstrate that a {\it star} graph becomes a {\it clique} 
when we double the linking length.
We refer to this as {\it star-clique transition} in spatial FOF networks. 
In more complex real-world networks, 
the transition may not be as clearly visible as Figure~\ref{fig:10} demonstrates. 
However, the transitional feature can be detected statistically 
in the centralization measurements 
{\it at the half scale} of the clique feature. 
The bottom-right panels in figures~\ref{fig:11} and~\ref{fig:12} show the centralization measurements 
of degree centrality. 
Both the degree centralization and largest clique size curves show similar ``knee'' features 
at scales of 40\arcsec and 70\arcsec~respectively. 
In contrast, the HOD mocks only show featureless linear trends. 
This indicates that the Bo\"otes LAEs have statistically more star-like configurations at 40\arcsec~
and the larger size of the largest clique at 70\arcsec~ than the HOD mocks, 
implying the {\it star-clique transition} in the network of Bo\"otes LAEs. 

We have found two interesting clues of the {\it star} 
and {\it clique} configurations at 40\arcsec~ and 70\arcsec~ respectively. 
Though not as strong as TR70, these two features 
imply that TR70 is due to the larger clique sizes of observed LAEs than the HOD mocks, 
which in turn may suggest an environmental factor in the formation and evolution 
of LAEs at $z \approx 2.67$.

\subsubsection{Model\#2 is marginally preferred over Model \#1}

Finally, we compare the differences between Model\#1 and Model\#2. 
In all 8 measurements,  Model\#2 shows better matches with the observed LAEs than Model\#1, 
though no special improvements can be found for explaining the three anomalies, 
CC170, TR70, and ED100, for Model\#2 either. 
The major improvements of Model\#2 from Model\#1 are marked using the solid blue triangles in Figure~\ref{fig:12}. 
The local statistics of average CC, transitivity, size of the largest clique, and degree centralization 
are larger in Model\#2 than in Model\#1 due to the higher fraction of central galaxy occupation, 
which we have referred to as ``Pristine Core Scenario''.  
These increased local statistics fit better the topology of observed LAEs. 

Hence, the network statistics prefer Model\#2 of the ``Pristine Core Scenario'' that, 
at $z \approx 2.67$, the central galaxies in massive halos, $> 10^{12} h^{-1} M_\odot$, still 
need to be less dusty to emit Ly$\alpha$ photons, 
potentially due to some replenishing channels of pristine gas such as the cold mode accretion 
(e.g., Kere\v{s} et al. 2005; Dekel \& Birnboim 2006; Kere\v{s} et al. 2009).

\section{Summary and Discussion}

We have investigated the spatial distribution of LAEs at $z \approx 2.67$, using the two-point correlation function 
and network statistics. From single power-law fits, we measure the correlation length, $r_0 = 4 h^{-1}$ Mpc, 
and bias, $b_{LAE} = 2.2^{+0.2}_{-0.1}$, consistent with previous studies of LAEs at similar redshifts.  
The power-law slopes are more uncertain and less consistent than the measured correlation lengths 
due to the clearly visible inflection point in the observed correlation function at small scales; 
i.e., where the one-halo term of subhalo statistics dominates. 
To obtain more accurate two-point statistics at these small scales reflecting the halo substructure, 
we need a larger survey volume containing better statistics on the small-scale separations (i.e., at $<10^Ó$).
Many current and future surveys will provide more accurate small-scale statistics  
so that we can investigate the scale-dependent features in two-point statistics beyond the single power-law interpretations.

From the HOD analysis, we have obtained two disparate, but degenerate, models, 
Model\#1 and Model\#2, which suggest different scenarios for the central galaxies 
for $> 10^{12} h^{-1}~ M_\odot$ halos at $z \approx 2.67$. 
This degeneracy is a byproduct of the inevitable tradeoff 
between flexibility and interpretability of parametric model, 
since the 6 fitting parameters of our HOD function lead to an overfit 
to the observed angular clustering, caused by over-flexible functional shapes. 
The LAE phenomenon may be a short-lived phase of galaxies, and it is possible 
that the HOD for this population of emission line galaxies needs to be more flexible 
than the models used to fit more continuum-luminous populations.
Due to this tradeoff between flexibility and interpretability, 
we need to accept all non-rejected HOD models as possible scenarios.

From the measurements of network statistics, we have found three distinct anomalies, 
TR70, ED100, and CC170, none of which are reproduced by the mocks 
constructed from the HOD models.   
The most conspicuous anomaly is TR70, 
which is a feature in the transitivity curve at a scale of 70\arcsec ($1.42 h^{-1}$~comoving Mpc).
From the additional measurements of 
{\it the size of largest clique} and {\it degree centralization}, we argue that TR70 reflects 
a strong environmental effect on forming LAEs within the diameter 
of $1.42 h^{-1}$ Mpc in the comoving scale and 570 kpc in the physical scale at $z \approx 2.67$.
The on-going and future spectroscopic surveys of LAEs, 
such as Hobby-Eberly Telescope Dark Energy Experiment (HETDEX; Hill et al. 2008), 
can provide definitive data sets for nailing down whether this environmental effect really exists 
and provide the redshift evolution of this {\it transitivity} peak.

Model\#2 works better for matching the graph topology of observed LAEs than Model\#1;  
especially the statistics of {\it average CC, transitivity, size of the largest cliques,} 
and {\it degree centralization} at small scales, $< 70$\arcsec.  
This suggests that the central halo occupation fraction of LAEs for massive halos 
should be large enough for generating more triangular and clique-like structures than the 
Dusty Core Scenario, Model\#1, predicts. 
Hence, at $z \approx 2.67$, the central galaxies in $> 10^{12} h^{-1}~ M_\odot$ halos  
need to be still less dusty to be bright enough in Ly$\alpha$ emission as LAEs, 
potentially due to some replenishing channels of pristine gas such as the cold mode accretion, 
along with appropriate geometrical vents, configured   
for unleashing Ly$\alpha$ photons from the star forming cores.

Statistics of network topology are more specialized in quantifying topological textures, 
while n-point statistics more specialized in quantifying geometric configurations.  
Although there are many reliable estimators of two- and three-point statistics 
for discrete observables; i.e., galaxy point distributions, 
n-point functions are intrinsically defined based on continuous observables; 
i.e., scalar fields such as cosmic density contrast and CMB temperature map. 

On the other hand, network statistics are inherently defined for quantifying discrete observables. 
Hence, at least in this perspective, graph analyses are more relevant 
for the investigation of spatial distributions of galaxies than n-point measurements.  
However, the inevitable weaknesses of {\it bias} and {\it shot noise} in galaxy distribution 
can affect graph statistics more directly than n-point statistics, since a couple of points 
can change the global pathways in galaxy network. 
We need, therefore, an ensemble of the discrete data   
to properly estimate how much such discrete impediments affect the overall graph measurements. 

These two kinds of statistics are complementary, 
since they quantify the galaxy point distribution from different perspectives. 
We can achieve unprecedentedly comprehensive views on 
galaxy distributions by measuring both of graph topology and n-point statistics, 
to precisely reveal evasive features of the matter distribution in the Universe.

\acknowledgments
We are grateful to the anonymous referee for comments that have improved this paper. 
SH's research activities have been supported by the University of Texas at Austin 
and Korea Institute for Advanced Study. 
Based in part on data collected at Subaru Telescope, which is operated by the National Astronomical Observatory of Japan. The observations reported here were obtained in part at the MMT Observatory, a facility operated jointly by the Smithsonian Institution and the University of Arizona. Some of the MMT telescope time was granted by NOAO, through the Telescope System Instrumentation Program (TSIP). TSIP is funded by NSF. 
AD's research is supported by the National Optical Astronomy Observatory, which is operated by the Association of Universities for Research in Astronomy (AURA) under cooperative agreement with the National Science Foundation. This research was also supported in part by NASA HST-GO-13000.

\appendix

\setcounter{figure}{0}
\makeatletter 
\renewcommand{\thefigure}{A\arabic{figure}}

\section{Empirical Random Pair Functions}\label{sec:appendixa}

\begin{figure}[t]
\centering
\includegraphics[height=5.0 in]{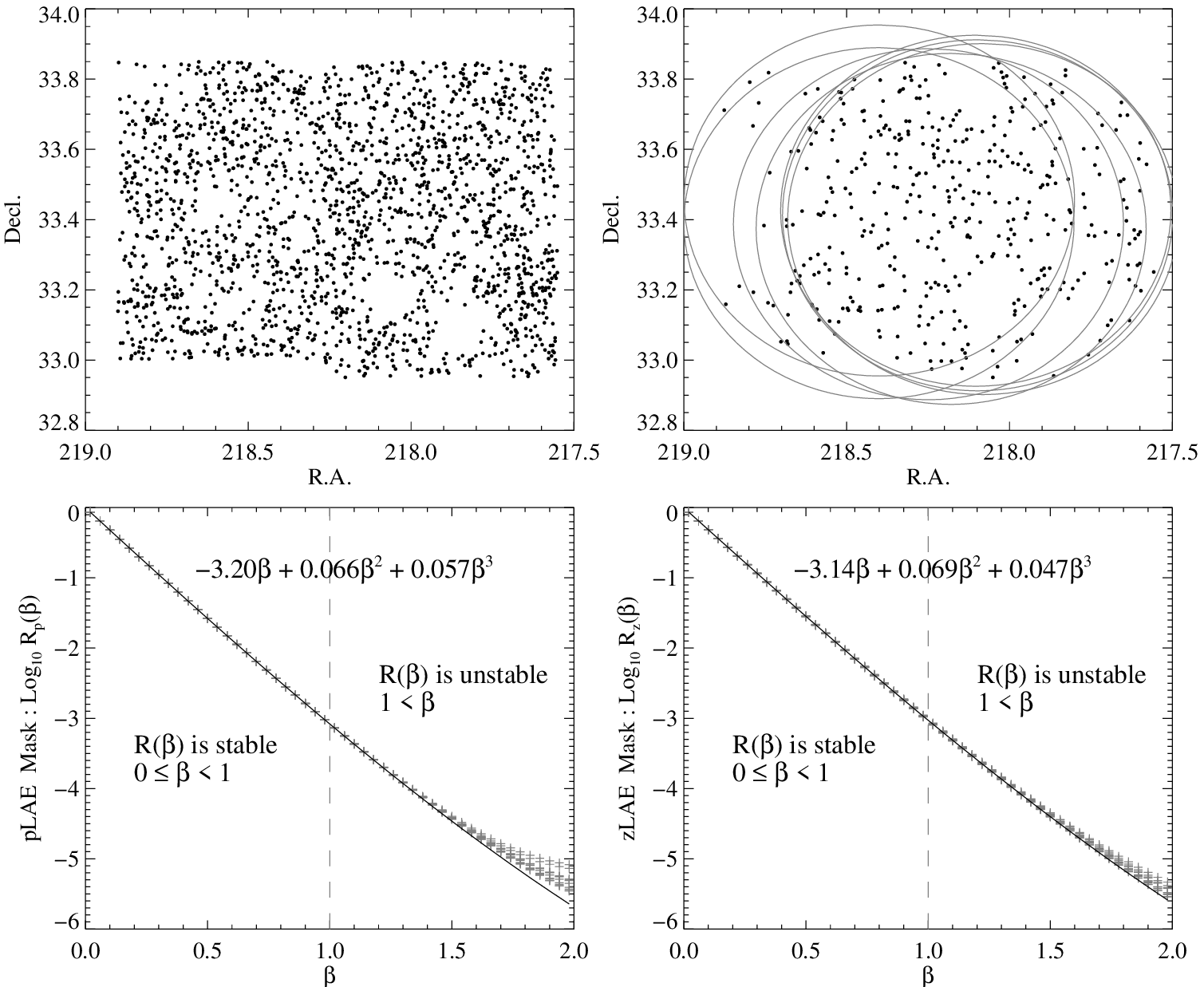}
\caption{The random sets for pLAE (top-left) and zLAE(top-right), 
and the corresponding RPFs, as defined in Equation~\ref{eq:rr}, 
for pLAE (bottom-left) and for zLAE (bottom-right). 
For $ 0 \le \beta < 1 $, the self-consistent fit in Equation~\ref{eq:rpf} is well defined mathematically. 
}\label{fig:4}
\end{figure}

Here, we describe the details about the divergence of Random Pair Functions (RPFs) 
and present their numerical forms, mentioned in \S\ref{sec:poweric}. 
First, we recall the definition of RPF : 
\begin{eqnarray}
R_\Omega (\beta) &\equiv& \frac{\sum RR~ \theta^{-\beta}}{\sum RR} \label{eq:rpfappendix}, 
\end{eqnarray}
where $RR$ is random pairs used for the LS estimator. 
From the definition, we can find two basic properties of RPF :  
(1) for a given $\beta$, RPF only depends on the random set, $RR$,    
and (2) for $\beta = 0$, $R_\Omega (\beta = 0) = 1$. 
The first property indicates that RPF depends only on the geometric shape of survey volume, 
like the geometric form factor of the LS estimator. The second property guarantees that 
RPF is, at least, well-defined at $\beta = 0$. 
For other $\beta$ values, it depends on the divergence of the integral sum,  $\int_{0}^{\infty} \theta^{-\beta} d\theta$, 
whether RPF is well defined or not.

The integral sum of  $\int_{0}^{\infty} \theta^{-\beta} d\theta$ is divided into three categories according to the values of $\beta$. 
For $0 \leq \beta < 1$,  the tail sum of  $\int_{1}^{\infty} \theta^{-\beta} d\theta$ diverges, 
while its local sum of  $\int_{0}^{1} \theta^{-\beta} d\theta$ is finite. We refer to this as ``large scale divergence''.  
Conversely, for $1 < \beta$, the local sum diverges, while the tail sum is finite. 
We refer to this as ``small scale divergence''.  
For $\beta = 1$, the sum diverges logarithmically in both small and large scales. 
Generally, since observational surveys cover finite portions of sky, 
RPFs are well-defined functions for  $0 \leq \beta < 1$; i.e., the sums in RPF are always finite real numbers.

Figure~\ref{fig:4} shows the two random sets for pLAE and zLAE (top panels) and their corresponding RPFs (bottom panels). 
The grey cross points show the RPFs for 10 different random sets and we fit them, in the range of  $0 \leq \beta < 1$, 
to obtain their numerical forms using cubic polynomials; 
\begin{eqnarray}
\log_{10} R_p (\beta) &=& -3.20\beta + 0.066\beta^2 + 0.057\beta^3 + O(\beta^4) \textrm{  for pLAE}\label{eq:fita}, \\
\log_{10} R_z (\beta) &=& -3.14\beta + 0.069\beta^2 + 0.047\beta^3 + O(\beta^4) \textrm{  for zLAE.} \label{eq:fitb}
\end{eqnarray}
The constant terms in Equation~\ref{eq:fita} and \ref{eq:fitb} are zero 
due to the boundary condition, $R_\Omega (\beta = 0) = 1$. We can find that the dominant terms are the first-order terms. 
The other higher order terms, $\beta^2, \beta^3, \cdots $, are minor and well truncated within  $0 \leq \beta < 1$. 

For $1 < \beta$, the higher order terms, $\beta^2, \beta^3, \cdots $, are divergent, rather than truncated. 
And a small fraction of very close pairs (i.e., $\theta << 1$) dominates the total sum in RPF. 
Hence, the RPF values become very unstable (having extreme variance) for random sets due to the ``small scale divergence''. 
Therefore, the self-consistent fit in Equation~\ref{eq:rpf}, 
\begin{eqnarray}
\omega_{LS}(\theta) &=& \frac{ \theta^{-\beta}  - R_\Omega (\beta) }{A_\omega^{-1}+ R_\Omega (\beta)}, \nonumber
\end{eqnarray}
is not valid for $1 < \beta$. 
Fortunately, the fiducial value for $\beta$ in most practical cases is near 0.8; hence, within $0 \leq \beta < 1$. 
Equation~\ref{eq:rpf} is, therefore, applicable in most cases.

\setcounter{figure}{0}
\makeatletter 
\renewcommand{\thefigure}{B\arabic{figure}}

\section{Physical Relevance vs. Statistical Interpretability in Parametric Models}\label{sec:appendixpara}

In this section we discuss which halo occupation distribution 
(HOD) reliably represents the halo occupation of Bo\"otes LAEs. 
We start with one of the most commonly used HODs,  
\begin{eqnarray}
N_c(M) &=& \left\{ \begin{array}{ll}
 0 & \textrm{if $M < M_c$},\\
 1 & \textrm{if $M \ge M_c$},
  \end{array} \right. \\ 
N_s(M) &=& \Big( \frac{M}{M_{1}} \Big)^\alpha, \label{eq:satz}
\end{eqnarray}
where $N_c(M)$ represents central galaxy distribution and $N_s(M)$ 
 satellite galaxy distribution for a given halo mass, $M$ 
(e.g., Zehavi et al. 2005; hereafter, we refer to this HOD as ZehaviHOD). 

\begin{figure}[t]
\centering
\includegraphics[height=2.5 in]{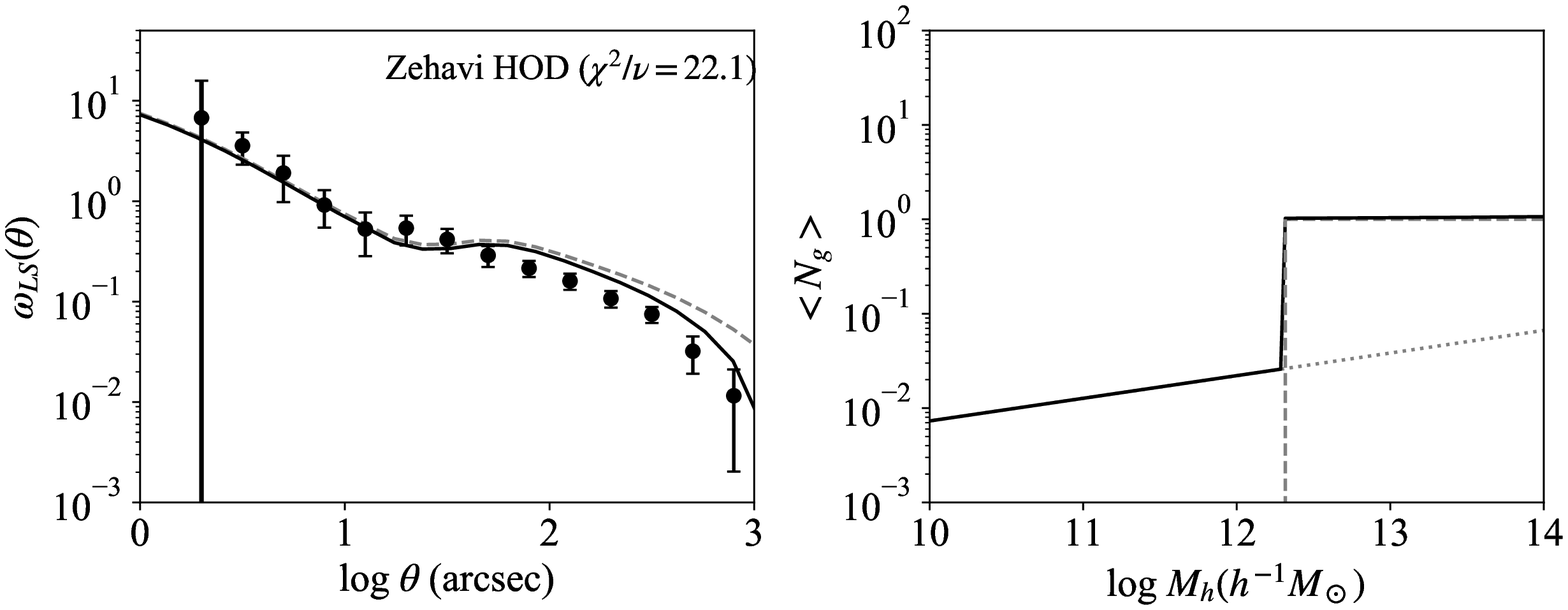}
\caption{The two point correlation functions (left) for observed LAEs (dots with error bars) 
and ZehaviHOD (lines; Zehavi et al. 2005), 
and halo occupation distribution (HODs; right) for ZehaviHOD. 
On the left panel, the dashed line represents the true two point function 
from the HOD Model without any effect of survey volume size, and the solid line the inversely 
corrected two point function using our inverse integral constraint method. 
On the right panel, the dashed line represents the average number of central galaxies, 
the dotted line of satellite galaxies, and the solid line of total galaxies. 
The HOD parameters for this model are $\log M_1 = 18.9$ and $\alpha = 0.24$, 
selected from the posterior probability function, shown in Figure~\ref{fig:b2}, 
with the estimates of  $\log M_1 = 17.7^{+1.6}_{-1.6}$ and $\alpha = 0.31^{+0.18}_{-0.08}$. 
}\label{fig:b1}
\end{figure}

\begin{figure}[t]
\centering
\includegraphics[height=3.0 in]{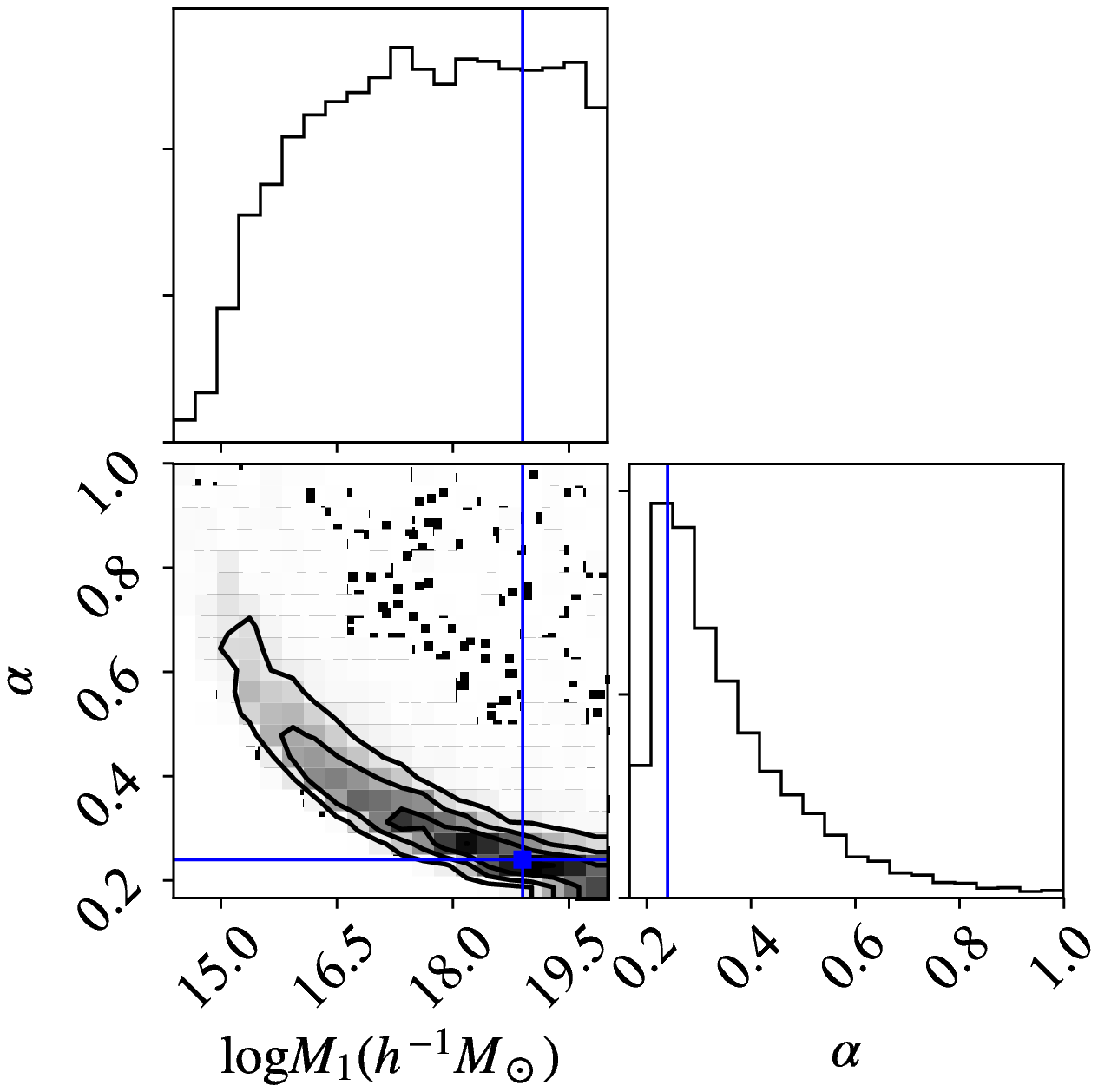}
\caption{The posterior probability function from the MCMC run for ZehaviHOD, visualized 
in contours (2D marginalized probabilities) and histograms (1D marginalized probabilities). 
We put 40 walkers (hence, 20 for each parameter) and run 700 steps. 
We discard the early 180 steps as burn-in and take 520 steps to retrieve the posterior probability function. 
The median values with $\pm 1\sigma$ errors for parameters are $\log M_1 = 17.7^{+1.6}_{-1.6}$ and $\alpha = 0.31^{+0.18}_{-0.08}$. 
The (blue) lines and point represent the location of our parameter choice of $\log M_1 = 18.9$ and $\alpha = 0.24$, 
where we demonstrate its two-point correlation function and HOD in Figure~\ref{fig:b1}. 
}\label{fig:b2}
\end{figure}

Figure~\ref{fig:b1} shows the two-point function (left) and HOD (right) for a ZehaviHOD,  
where we choose its model parameters from the posterior probability function 
shown in Figure~\ref{fig:b2}, obtained using MCMC sampling. 
All results shown in Figure~\ref{fig:b1} and \ref{fig:b2} are rough estimates, 
since the current outputs are already unlikely, $\log M_1 = 17.7^{+1.6}_{-1.6}$  
and $\alpha = 0.31^{+0.18}_{-0.08}$, indicating that ZehaviHOD is {\it not physically relevant} 
for describing emission line galaxies (ELGs). 
For example, the current ZehaviHOD predicts that 
the small halos with masses of $< 10^{12}$ $M_\odot$ do not host LAEs at their centers; 
if any, they should be satellites. For massive halos, even if they host dust obscured galaxies at their centers, 
they should be detected in Ly$\alpha$ emission. 

The {\it duty cycle} of LAEs is one of the main reasons why ZehaviHOD fails to be a relevant model. 
Unlike red dwarf stars serve as lifelong emission sources for galaxy, 
Ly$\alpha$ emission is only lit up for a short period of time. 
Hence, the halo occupation fraction should be allowed to be smaller than one. 
We write down a new HOD using duty cycles as, 
\begin{eqnarray}
N_c(M) &=& \left\{ \begin{array}{ll}
 0 & \textrm{if $M < M_c$},\\
 F_c & \textrm{if $M \geq M_c $},
  \end{array} \right.  \label{eq:offa} \\ 
N_s(M) &=& F_s\Big( \frac{M}{M_{1}} \Big)^\alpha, \label{eq:off} 
\end{eqnarray}
where $F_c$ is a duty cycle for central LAEs and $F_s$ for satellite LAEs. 
By {\it adding these two new parameters}, we can achieve {\it more physically relevant} predictions 
to the halo occupation of LAEs. However, as a trade-off, we can lose statistical interpretability 
due to coupled parameters; moreover, there exists a potential degeneracy in model fits.  

For example, 
the prediction of $\alpha \approx 0.3$ from ZehaviHOD is quite smaller than a fiducial value, $\alpha = 1$. 
This is because the one-halo term is determined by center-satellite and satellite-satellite pair counts. 
Since the central occupation fraction is always equal to one for massive halos in ZehaviHOD, 
the number of satellites should be suppressed by  taking unphysically high $\log M_1 \approx 17$ and low $\alpha$ 
to match the observed small-scale clustering. 
If we take a small $F_c$, we can have a more parametric freedom to increase the number of satellites, 
while still fixing the total pair counts of center-satellite and satellite-satellite. 
Hence, by adding {\it duty cycles} to our new HOD, 
we can achieve more physically relevant predictions to $\log M_1$ and $\alpha$. 
 
However, as a trade-off, we have a coupled factor, $F_s M_1^{-\alpha}$, 
for the satellite occupation in Equation~\ref{eq:off}. 
Though fixing this factor a constant, 
there are internal degenerate degrees of freedom among $\{ F_s, M_1, \alpha \}$. 
In addition, $\{ F_c \}$ is coupled with 
the satellite occupation parameters $\{ F_s, M_1, \alpha \}$, 
which determines the number of center-satellite pairs, affecting small-scale clustering significantly. 

Therefore, we can obtain a better HOD model by increasing its flexibility of functional form.  
However, we lose the model's interpretability due to explicit and implicit couplings among parameters 
and potentially the  degeneracy increases in parameter estimates. 
If the  duty cycle of LAEs is inevitably required for  physical relevance, 
its related trade-offs are intrinsically ineluctable.

Before taking Equation~\ref{eq:offa} and \ref{eq:off} as our final HOD choice, 
we need to consider one more factor, the  environmental effect on populating central LAEs. 
The question is whether it is physically relevant to populate the same fraction of LAEs at centers 
for different halos in various environments; for example, $10^{11} M_\odot$ halos mostly 
populated in field regions and $10^{13} M_\odot$ in dense regions. 
In a practical aspect,  we need to decide whether it is necessary to add another set of parameters 
to the LAE's HOD for implementing such mass-dependent occupations.  
Unlike the  duty cycle, this could be arguably optional 
for  physical relevance, when considering 
the caveats of additional trade-offs caused by the new parameters. 

For our sample, we have a conspicuous inflection point near 20\arcsec, 
which implies that the substructures of massive halos, determining the small-scale clustering, 
should be more accurately treated to properly explain the inflected feature. 
Therefore, we assign two different fractions of central occupations for the halos as, 
\begin{eqnarray}
N_c(M) &=& \left\{ \begin{array}{ll}
 0 & \textrm{if $M < M_c$},\\
 F_c^B & \textrm{if $M_c \leq M \leq M_\theta $},\\
 F_c^A & \textrm{if $M_\theta < M $}, \label{eq:b5}
  \end{array} \right. 
\end{eqnarray}
where $F_c^A$ and $F_c^B$ are central occupation fractions, 
split by a mass threshold $M_\theta$. 
Using this equation, we can assign different central occupations, for example, 
to $10^{11} M_\odot$ and $10^{13} M_\odot$ halos. 
As trade-offs, we have an explicit coupling among $\{ F^A_c, F^B_c, M_\theta \}$
and an implicit dependence between $\{ F^A_c, F^B_c, M_\theta \}$ and $\{ F_s, M_1, \alpha \}$. 
Despite the issues of poor  interpretability and potential  degeneracy, we argue that 
the central occupations of LAEs for  $10^{11} M_\odot$ and $10^{13} M_\odot$ halos 
should be different at $z \approx 2.67$. To conclude, by implementing the two physical factors of 
(1)  duty cycles and (2)   mass-dependent central occupations, 
our choice of physically relevant HOD for LAEs is Equation~\ref{eq:off} and~\ref{eq:b5} 
with the 6 parameters $\{ F_c^A,  F_c^B, F_s, M_\theta, \alpha, M_1 \}$. 

In the literature, Geach et al. (2012) already implemented the two physical factors as, 
\begin{eqnarray}
N_c(M) &=& F_c^B(1-F_c^A) \exp\Big[ - \frac{\log(M/M_c) ^2}{2 \sigma^2_{\log M}} \Big] \nonumber \\ 
&+& F_c^A\Big[ 1 + \textrm{erf} \Big(\frac{\log(M/M_c)}{ \sigma_{\log M} } \Big) \Big], \label{eq:cen} \\ 
N_s(M) &=& F_s\Big[ 1 + \textrm{erf} \Big(\frac{\log(M/M_{1})}{ \delta_{\log M} } \Big) \Big] \Big( \frac{M}{M_{1}} \Big)^\alpha. 
\end{eqnarray}
where the main difference from Equation~\ref{eq:off} and~\ref{eq:b5} is a smoother mass-dependence using Gaussian distribution 
with one additional parameter, $\delta_{\log M}$. 
When fixing $\delta_{\log M} \equiv 1$, the parameter set of Geach et al. is $\{ F_c^A,  F_c^B, F_s, \sigma_{\log M}, \alpha, M_1 \}$, 
while $\{ F_c^A,  F_c^B, F_s, M_\theta, \alpha, M_1 \}$ of Equation~\ref{eq:off} and~\ref{eq:b5}. 
Therefore, we adopt the HOD from Geach et al. for the Bo\"otes LAEs for  physical relevance
considering the two factors of  duty cycles and  mass-dependent central occupations. 
Due to the inevitable trade-offs of poor  interpretability and potential  degeneracy, 
we accept all non-rejected HOD models as possible scenarios.

\setcounter{figure}{0}
\makeatletter 
\renewcommand{\thefigure}{C\arabic{figure}}

\section{Definitions of Network Quantities}\label{sec:appendixb}

All graph quantities presented in this section are commonly used in network science. 
Interested readers are referred to Newman (2003), 
Dorogovtsev, Goltsev \& Mendes (2008), and Barth\'elemy (2011) for further details.

The {\it Average Clustering Coefficient} (average CC) is an average of all local clustering coefficients. 
The local clustering coefficient $C_i$ for a vertex $i$ is defined as, 
\begin{eqnarray}
C_i & = & \frac{\textrm{number of pairs of neighbors for }i ~\textrm{that are connected}}
{\textrm{number of pairs of neighbors for }i}.\label{eq:acc}
\end{eqnarray} 
In social networks, the local clustering coefficient measures whether an individualÕs two friends know each other.
The denominator in Equation~\ref{eq:acc} is the number of total pair combinations of the individual's friends. 
The numerator is the number of friended pairs; hence, triangular friendships when including the central individual. 
The local clustering coefficient, therefore, is roughly a triangle density for each vertex. 
The average of this vertex-wise triangle density is the average CC for a network. 

\begin{figure}[t]
\centering
\includegraphics[height=3.0 in]{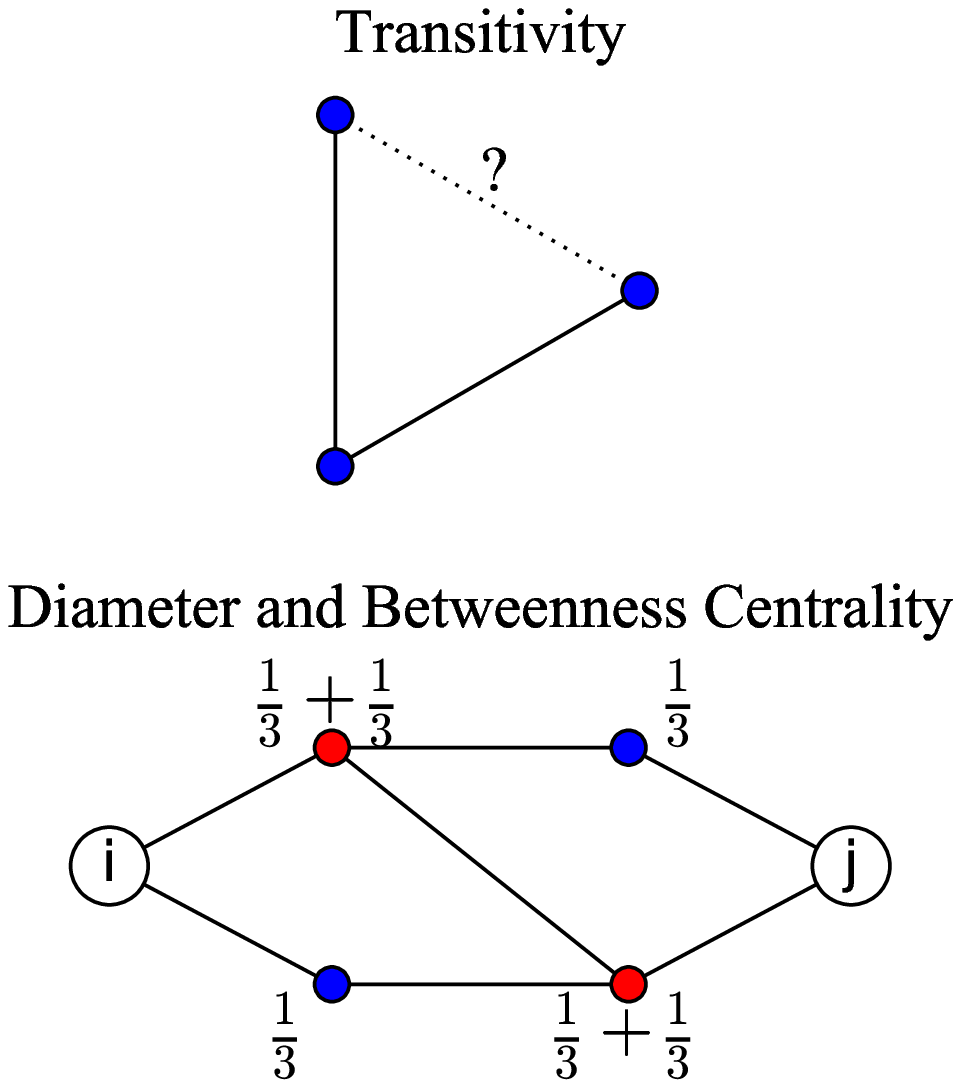}
\caption{The schematic figure illustrating the meanings of transitivity, diameter, and betweenness centrality. 
}\label{fig:8}
\end{figure}

{\it Transitivity} is a different version of triangle density from the average CC, defined as:  
\begin{eqnarray}
\textrm{Transitivity} & = & \frac{3 \times \textrm{number of triangles}}{\textrm{number of connected triples}}.\label{eq:tr}
\end{eqnarray}
The top graph schema in Figure~\ref{fig:8} illustrates the meaning of transitivity. 
The ``$\vee$'' configuration, connected by solid lines, is a {\it connected triple}. 
Transitivity is the fraction of whether the other side, drawn by a dotted line, is connected or not. 
Since a triangle contains three connected triples, transitivity is normalized to 1 as the average CC. 
Transitivity is often referred to as global clustering coefficient, 
since Equation~\ref{eq:tr} is a network-wise measurement while Equation~\ref{eq:acc} a vertex-wise measurement. 
Hence, we need to measure transitivity for a true unbiased triangle density for a network. 
The average CC is biased to the majority's CC value in vertex population due to the averaging process. 
Therefore, transitivity and average CC are similar, but not exactly the same. 

\begin{figure}[t]
\centering
\includegraphics[height=1.0 in]{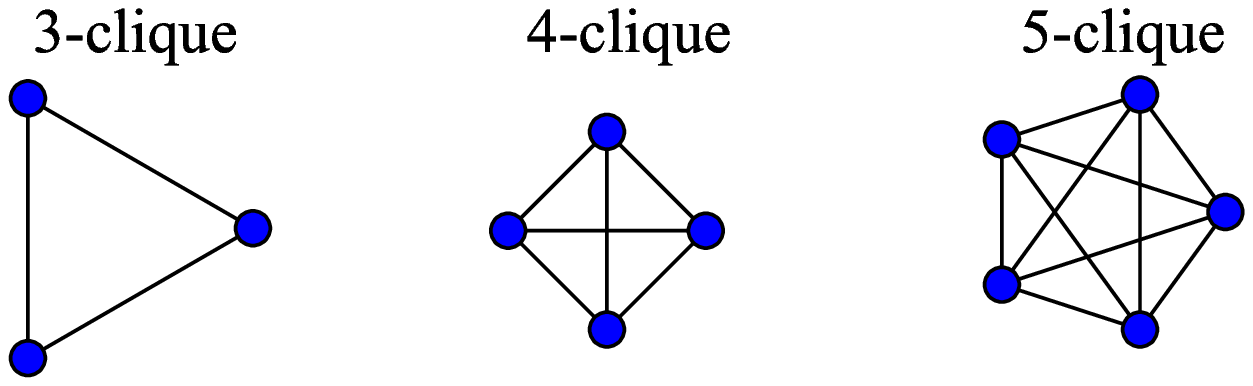}
\caption{A schematic figure showing 3, 4, and 5-cliques, where we refer to a complete subgraph with k--vertices as {\it k--clique}. 
Inside of the 5-clique, we can find many 3-- and 4-- subcliques. Generally, we can extend a clique by adding neighbors, 
until there is no more extendable clique configuration. This kind of {\it un-extendable} clique is referred to as {\it maximal clique}. 
Since galaxy groups and clusters form cliques in galaxy FOF networks, 
statistics of maximal cliques are quite interesting and important information 
for investigating the formation and evolution of galaxy groups and clusters.   
}\label{fig:9}
\end{figure}

A {\it clique} is a complete subgraph. Figure~\ref{fig:9} show cliques with 3,4, and 5 vertices; 
hereafter, we refer to a clique with $k$ vertices as {\it k-clique}. 
Inside of the 5-clique, we can find many 3- and 4-subcliques. Generally, we can extend a clique 
by adding neighbors, until there is no more extendable clique configuration. 
This kind of un-extendable clique is referred to as {\it maximal clique}. 
Since galaxy groups and clusters form cliques in galaxy FOF networks, 
statistics of maximal cliques are quite interesting and important information 
for investigating the formation and evolution of galaxy groups and clusters. 
We find the largest maximal clique and measure its size from each network. 

The {\it Diameter} is the largest path length of shortest pathways from all pairs in a network. 
The path length is defined as the number of steps to reach from a certain vertex, $i$, to another, $j$. 
Hence, the pathways of minimum path length are the shortest pathways between the vertices, $i$ and $j$; 
generally, there can be multiple shortest pathways between a pair in an unweighted network. 
The bottom graph schema in Figure~\ref{fig:8} illustrates the shortest pathways between $i$ and $j$ vertices. 
There are three shortest pathways with the path length of 3. 
And there is one detour with the path length of 5. Therefore, the shortest path length between $i$ and $j$ 
is 3. We measure these shortest path lengths for all possible pairs in a network and, then, take the maximum value. 
This largest path length is defined as the {\it diameter} of the network. 

{\it Centrality} is a value assigned to each vertex, as an indicator for quantifying which vertex 
is more important in a certain topological perspective. 
For example, {\it Degree Centrality} is the number of neighbors for each vertex. 
In social networks, this is a measure of the importance of a given individual in the network; the most influential individual is the one with the most ``friends", i.e., the one with the highest degree value. 
A better centrality can be defined 
if the current centrality cannot reflect the concerned topological feature well. 
The Google's {\it PageRank} is designed to prioritize the importance of World-Wide Web (WWW) documents. 
This centrality works better to rank WWW documents than the simple degree centrality (Page et al. 1999). 

The {\it Betweenness Centrality} is a measure of which  vertex is most frequently used when commuting back and forth 
between all pairs; hence, the congested spots during rush hours have high betweenness centralities in a road network. 
Mathematically, this betweenness, $x_i$ for the $i$-th vertex, is defined as:
\begin{equation}
x_i = \sum\limits_{st} \frac{n^i_{st}}{g_{st}}, \label{eq:betweenness}
\end{equation}
where $g_{st}$ is the number of shortest paths between the vertices $s$ and $t$, 
and $n^i_{st}$ the number of these which pass through the vertex, $i$. 
If $g_{st}$ is zero, we assign $n^i_{st}/g_{st} = 0$. 
In the bottom graph schema of Figure~\ref{fig:8}, there are 3 shortest pathways between $i$ and $j$. 
By the definition of betweenness in Equation~\ref{eq:betweenness}, we add $+1/3$ to all vertices 
on each shortest pathway. Then, $+2/3$ is assigned to red vertices and $+1/3$ is assigned to blue vertices by the pair of $i$ and $j$. 
We cumulate all of these betweenness values from all pairs to obtain the final betweenness centrality. 
Generally, this betweenness can be used to identify which spot is the most congested area in a road network or 
which person is the most influential broker connecting two isolated communities. 
In galaxy FOF networks, betweenness can be used as a {\it filament tracer} (Hong \& Dey 2015). 

\begin{figure}[t]
\centering
\includegraphics[height=2.0 in]{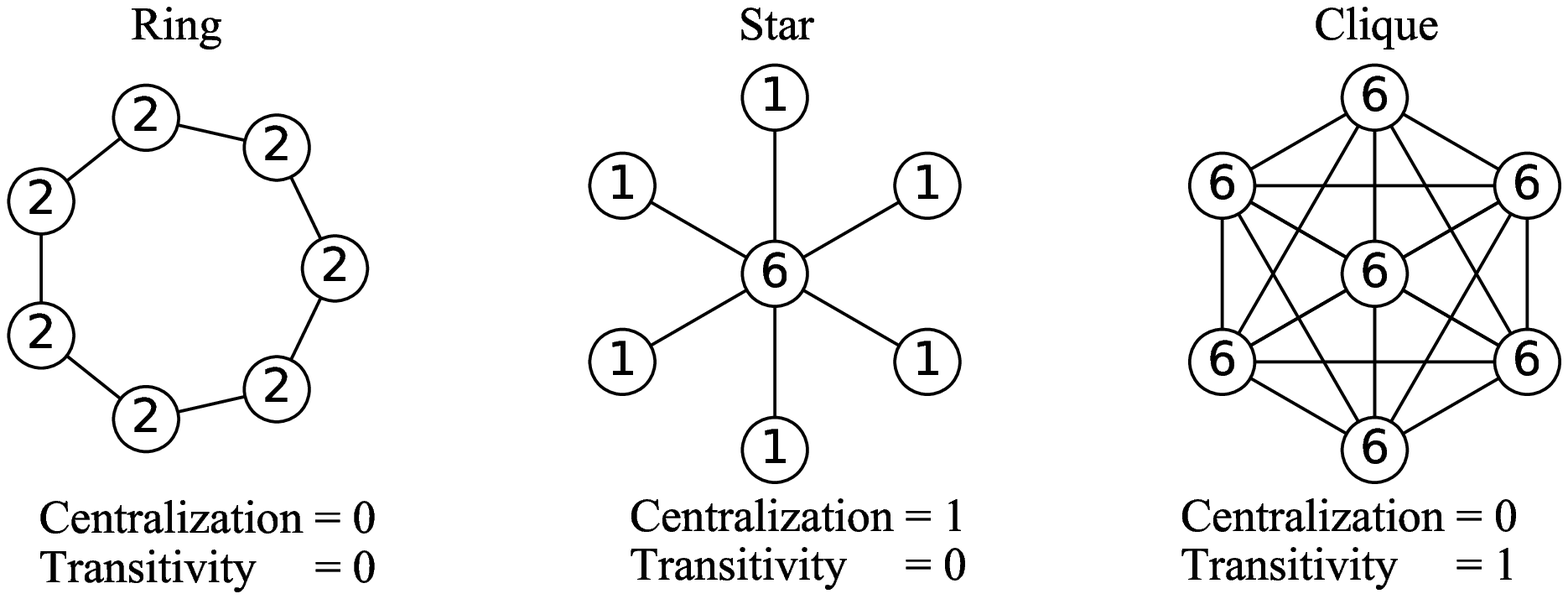}
\caption{A schematic figure showing Ring (left), Star (middle), and Clique (right) graphs with 7 vertices. 
The number on each vertex represents the number of neighbors (i.e., {\it degree}), 
and {\it centralization} and {\it transitivity} values are shown at the bottom for each graph. 
It is interesting that, in our FOF recipe for generating networks, a star graph becomes a clique when the linking length is doubled; 
hereafter, we refer to this as {\it star--clique transition}.  
Figures~\ref{fig:11} and \ref{fig:12} provide some potential evidence of this star--clique transition. 
}\label{fig:10}
\end{figure}

Like the local clustering coefficient, betweenness and degree are vertex-wise measurements. 
As we average out local clustering coefficients to an average CC, we can measure 
the averages of betweenness and degree. However, for centralities, there is another way 
of reducing the vertex-wise values, 
referred to as {\it centralization}, which quantify how close a network is to a star graph, 
the most centralized graph structure. 
There are a couple of ways to define centralization. In this paper, we follow the Freeman's formula, 
\begin{equation}
\textrm{Centralization} = \frac{\sum\limits_{i=1}^{n} [ C_{max} - C_i ]}{(n-1)(n-2)},
\end{equation}
where $C_i$ is a centrality value for a vertex $i$ and $C_{max}$ the maximum value of centrality. 
Figure ~\ref{fig:10} shows Ring (left), Star (middle), and Clique (right) graphs with 7 vertices. 
The number on each vertex represents the number of neighbors (i.e., degree centrality), 
and the corresponding degree centralization and transitivity values are shown at the bottom of each graph. 
This demonstrates well how we can quantitatively discern the different kinds of network configurations 
using centralization and transitivity. 
In galaxy FOF networks, there is an interesting connection between star and clique   
that a star graph becomes a clique when we double the linking length from where a star graph forms. 
We refer to this as {\it star--clique transition}. If we find some anomaly in clique statistics, 
we may expect some related abnormal feature in centralization statistics at the half scale 
from where we find the clique anomaly.

The {\it Giant component} is the largest connected subgraph in a network. 
The giant components are trivial for the two extreme linking lengths in a galaxy FOF network.  
For a small linking length that isolates all individual galaxies, the size of the giant component 
is trivially 1. 
In the opposite case of a very large linking length forming a complete graph, 
the giant component size is equal to the total number of vertices (galaxies). 
Hence, the ratio of the size of giant component to the total number of vertices 
is a fraction that increases from 0 to 1 monotonically as the linking length grows from zero. 
This growth rate of giant component fraction depends on topology; 
especially aligned bridging structures like filaments, 
which connect vertices more efficiently than featureless random scatters. 
In this case, the fraction of giant component grows faster through the bridges 
to reach 1 at a smaller linking length than in the case of networks without such topological shortcuts.  

Finally, the {\it Edge Density} is the number of edges divided by the total number of possible pairs, $n(n-1)/2$, 
to be normalized to 1.

\end{document}